\documentclass[journal]{IEEEtran}
\usepackage{amsmath,amsfonts}
\usepackage{algorithmic}
\usepackage{algorithm}
\usepackage{array}
\usepackage[caption=false,font=normalsize,labelfont=sf,textfont=sf]{subfig}
\usepackage{textcomp}
\usepackage{stfloats}
\usepackage{url}
\usepackage{verbatim}
\usepackage{graphicx}
\usepackage{cite}
\usepackage{microtype} % for professional tables
\usepackage{hyperref}

\usepackage{amsthm}

\usepackage{algorithm}
\usepackage{algorithmic}
\usepackage{color}

\usepackage{multirow}
\usepackage{dcolumn}

\usepackage{xcolor}

\usepackage{url}
\usepackage{soul}

\usepackage{enumitem}
\setlist[itemize]{leftmargin=*}

\usepackage{xspace}

\usepackage{colortbl}

\usepackage{hyperref}
\hypersetup{
    colorlinks,
    linkcolor={black},
    citecolor={black},
    urlcolor={blue},
    pdfborder={0 0 0},
}

\usepackage{amsthm}

\usepackage{algorithm}
\usepackage{algorithmic}
\usepackage{color}

\usepackage{multirow}
\usepackage{dcolumn}

\usepackage{xcolor}

\usepackage{url}
\usepackage{soul}

\usepackage{enumitem}
\setlist[itemize]{leftmargin=*}

\usepackage{xspace}

\usepackage{colortbl}

\newcommand{\mypara}[1]{\smallskip\noindent\textbf{#1.} \xspace}

\definecolor{lightgray}{gray}{0.92}

\newtheorem{theorem}{\bf Theorem}
\newcommand{\Name}{\texttt{InferDPT}\xspace}
\floatname{algorithm}{Protocol}

\usepackage{booktabs} 
\newtheorem{definition}{\bf{Definition}}

\definecolor{RawSienna}{RGB}{190,0,0}
\definecolor{Blue}{RGB}{0, 51, 153}

\renewcommand{\epsilon}{\varepsilon}

\definecolor{light_red}{rgb}{0.96, 0.76, 0.76}
\definecolor{light_green}{rgb}{0.84, 0.94, 0.77}
\usepackage{soul}   

\newcommand{\bitone}[1]{\sethlcolor{light_green}\hl{#1}}

\begin{document}

\title{\Name: Privacy-preserving Inference for Closed-box Large Language Models}

\author{\IEEEauthorblockN{Meng Tong, %\IEEEauthorrefmark{1},
Kejiang Chen, %\IEEEauthorrefmark{1},
Jie Zhang, %\IEEEauthorrefmark{1}, 
Yuang Qi,\\
Weiming Zhang, %\IEEEauthorrefmark{1}, 
% Yaofei Wang, %\IEEEauthorrefmark{1},and 
% Yaofei Wang is affiliated with the School of Computer Science and Information Engineering at Hefei University of Technology.
Nenghai Yu,
Tianwei Zhang,
Zhikun Zhang
}%\IEEEauthorrefmark{1},}\\
% \IEEEauthorblockA{\IEEEauthorrefmark{1}School of Information Science and Technology, University of Science and Technology of China,
 % \\Key Laboratory of Electromagnetic Space Information, Chinese Academy of Sciences.\\}
% ~\IEEEmembership{Fellow,~IEEE}
% \IEEEauthorblockA{\IEEEauthorrefmark{2}Twentieth Century Fox, Springfield, USA}
% \IEEEauthorblockA{\IEEEauthorrefmark{3}Starfleet Academy, San Francisco, CA 96678 USA}
% \IEEEauthorblockA{\IEEEauthorrefmark{4}Tyrell Inc., 123 Replicant Street, Los Angeles, CA 90210 USA}% <-this % stops an unwanted space
%\thanks{This work was supported in part by the Natural Science Foundation of China under Grant U1636201, Anhui Initiative in Quantum Information Technologies under Grant AHY150400, Anhui Science Foundation of China under Grant 2008085QF296, and Exploration Fund Project of University of Science and Technology of China under Grant YD3480002001.}
\thanks{This work was supported in part by the National Natural Science Foundation of China under Grant U2436601, 62472398, U2336206, and 62121002 , and by the Open Foundation of Key Laboratory of Cyberspace Security, Ministry of Education (No.KLCS20240207), as well as by the National Research Foundation, Singapore and Infocomm Media Development Authority under its Trust Tech Funding Initiative (No. DTCRGC-04). Any opinions, findings and conclusions or recommendations expressed in this material are those of the author(s) and do not reflect the views of National Research Foundation, Singapore and Infocomm Media Development Authority.}
\thanks{M. Tong and K. Chen are with the University of Science and Technology of China and the Key Laboratory of Cyberspace Security, Ministry of Education, China. Y. Qi, W. Zhang, and N. Yu are with the University of Science and Technology of China.
J. Zhang is with CFAR, A*STAR.
T. Zhang is with Nanyang Technological University.
Z. Zhang is with Zhejiang University.}
\thanks{Corresponding authors: Kejiang Chen (Email:chenkj@ustc.edu.cn) and Weiming Zhang (Email: zhangwm@ustc.edu.cn)}
}

\markboth{IEEE Transactions on Dependable and Secure Computing}%
{Shell \MakeLowercase{\textit{et al.}}: Bare Demo of IEEEtran.cls for IEEE Transactions on Magnetics Journals}

\IEEEtitleabstractindextext{
\begin{abstract}
\textit{Large language models} (LLMs), represented by ChatGPT, have greatly simplified text generation tasks. However, they have also raised concerns about privacy risks such as data leakage and unauthorized information collection. Existing solutions for privacy-preserving inference face practical challenges related to computational time and communication costs. 
In this paper, we propose \Name, the first practical framework for privacy-preserving \underline{\texttt{Infer}}ence of black-box LLMs, implementing \underline{\texttt{D}}ifferential \underline{\texttt{P}}rivacy in \underline{\texttt{T}}ext generation. \Name comprises two key modules: the ``perturbation module" utilizes the differentially private mechanism to generate a perturbed prompt, facilitating privacy-preserving inference with black-box LLMs; the ``extraction module", inspired by knowledge distillation and phenomenon we observed, extracts coherent and consistent text from the perturbed generation result, ensuring successful text generation completion. 
%
%To address privacy concerns related to previous differentially private mechanisms' susceptibility to embedding inversion attacks
To achieve a better balance between utility and privacy protection, we introduce RANTEXT, a novel differentially private mechanism integrated into the perturbation module of \Name, which introduces the concept of ``\underline{RAN}dom adjacency list" for \underline{TEXT} perturbation within the prompt. 
Experimental results across three datasets demonstrate that the text generation quality of \Name is comparable to that of non-private GPT-4, and RANTEXT surpasses existing state-of-the-art mechanisms, namely, SANTEXT+ and CUSTEXT+ in the trade-off between privacy and utility. Even with a privacy parameter \(\epsilon\) value of 6.0, RANTEXT achieves an average privacy protection level of exceeding 0.90 against the embedding inversion attacks, which is 0.58$\times$ higher than that of SANTEXT+ and 3.35$\times$ higher than that of CUSTEXT+. Our code is available at: \href{https://github.com/mengtong0110/InferDPT}{https://github.com/mengtong0110/InferDPT}.
\end{abstract}
}

\maketitle

\IEEEdisplaynontitleabstractindextext

\IEEEpeerreviewmaketitle
\begin{IEEEkeywords}
Differential privacy, black box, inference, large language model.
\end{IEEEkeywords}
\setcounter{page}{1}
\section{Introduction}
\label{intro}

\IEEEPARstart{I}{n} recent years, the rapid advancement of \textit{large language models} (LLMs) has garnered widespread attention from both the academic and industrial communities worldwide~\cite{li-etal-2023-large}. %\zzk{More citations here}.
% particularly in natural language processing (NLP). 
ChatGPT~\cite{ChatGPT}, a prominent example, has reached a remarkable milestone with 100 million weekly active users, as announced by OpenAI CEO Sam Altman on November 6, 2023, during the company's inaugural developer conference held in San Francisco~\cite{his}.
The widespread popularity of ChatGPT has significantly facilitated people's daily work and lives. Users interact with ChatGPT via APIs or web interfaces to generate text for various applications, including but not limited to drafting articles, documenting daily work activities, and crafting advertisements for new products~\cite{zhao2023survey}.
% For example, it assists in drafting academic papers, documenting work, and even preparing introductions for upcoming commercial products.

\begin{figure*}[t]
\centering
\includegraphics[width=0.89\textwidth]{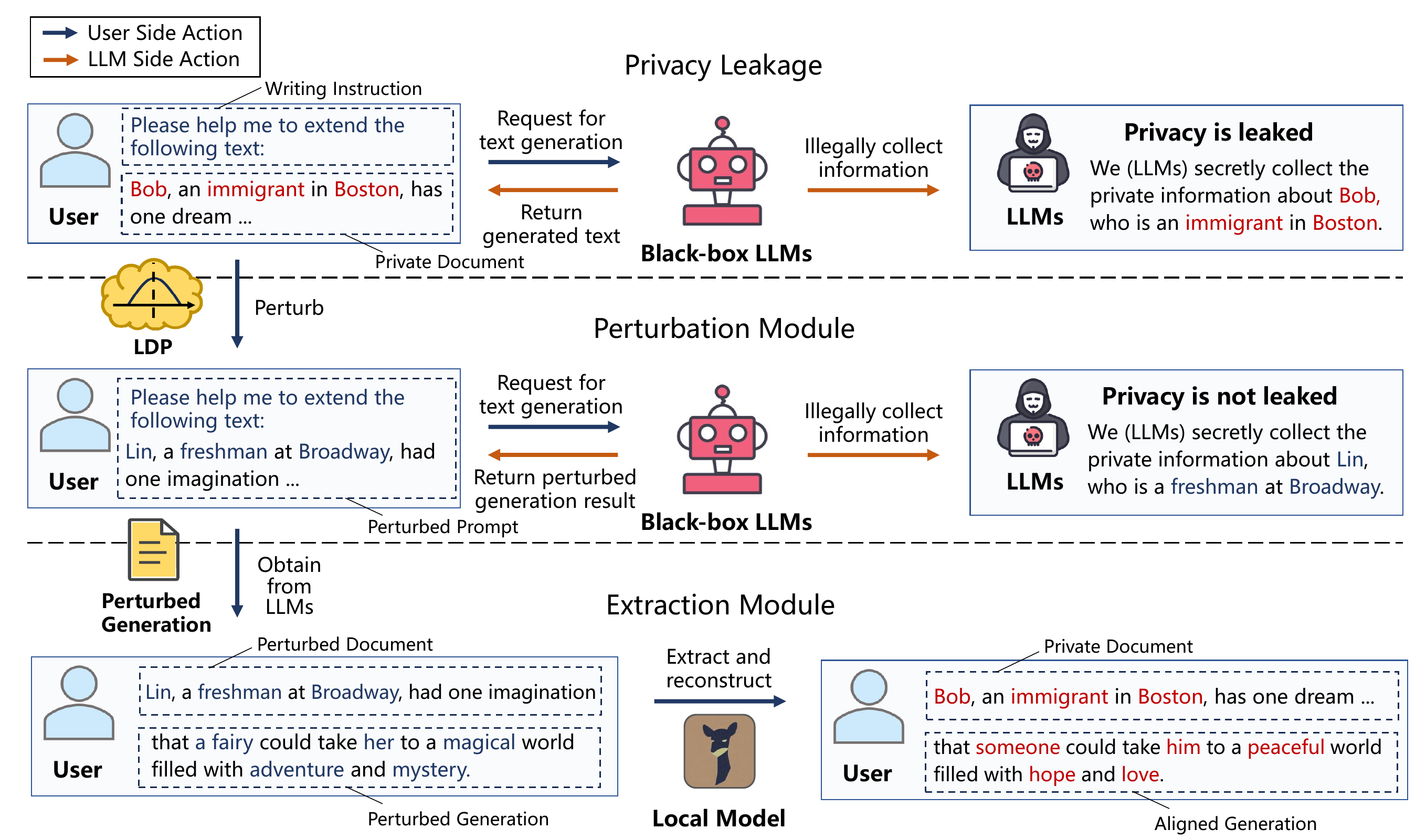}
\caption{The illustration of potential privacy leakage and a solution via \Name when a user employs black-box LLMs for text generation tasks.}
\label{fig:problem}
\end{figure*}

However, technology is a double-edged sword. 
While LLMs offer unparalleled convenience and utility in text generation, they may also raise potential privacy concerns.
There are instances where the misuse of LLMs has led to serious privacy infringements. 
One such example involves Samsung employees leaking the company's confidential meeting records and sensitive data about unreleased products~\cite{sam}.
Furthermore, in a recent incident, GPT-3.5 unexpectedly disclosed an individual's selfies~\cite{gpt3.5}. 
These incidents reignited concerns among the public regarding the potential privacy risks associated with uploading personal data to LLMs~\cite{Italy,claudeh}. 
Therefore, it is crucial to address privacy concerns of uploading query contents, which is called \textit{prompt}. We provide an example in \autoref{fig:problem} to demonstrate privacy leakage in the \textit{prompt} when a user interacts with LLMs.
% Generally, a \textit{prompt} in text generation tasks , as shown in \autoref{fig:problem}. 

% during the black-box inference in text generation with advanced LLMs.

% Specifically, employees of Samsung leaked the company's meeting records and critical data about unreleased products~\cite{sam}. In a recent incident, GPT-3.5~\cite{gpt3.5} unexpectedly disclosed an individual's selfies, which once again evoked the crowd's privacy concerns related to uploading private data to LLMs~\cite{Italy}.

\begin{table}[!t]
\centering
\caption{Comparisons of different methods. A check mark ({\color{red!80}\checkmark}) indicates that methods meet the scenario requirements.
%\zzk{Change the order in the table to be consistent with the text.}
}
\label{tab:comparison}
\resizebox{\columnwidth}{!}{%
\begin{tabular}{c|cccc}
\toprule
\textbf{Method} & \textbf{Text Generation}  & \textbf{Black Box}  & \textbf{Inference} & \textbf{Low Cost} \\
\midrule
CipherGPT~\cite{hou2023ciphergpt} & {} &{\color{red!80}\checkmark} &{\color{red!80}\checkmark} &\\
TextObfuscator~\cite{zhou2023textobfuscator} & &  &{\color{red!80}\checkmark} &{\color{red!80}\checkmark}\\
DP-Forward~\cite{du2023dp} & & &{\color{red!80}\checkmark} &{\color{red!80}\checkmark} \\
SANTEXT+~\cite{yue-etal-2021-differential} & & {\color{red!80}\checkmark} &  &{\color{red!80}\checkmark} \\
CUSTEXT+~\cite{chen2023customized} & & {\color{red!80}\checkmark} & &{\color{red!80}\checkmark}\\
\Name\!+ RANTEXT & {\color{red!80}\checkmark} & {\color{red!80}\checkmark} & {\color{red!80}\checkmark} &{\color{red!80}\checkmark}\\
\bottomrule
\end{tabular}%
}
\end{table}
\mypara{Existing Solutions}{A \textit{prompt} in text generation tasks consists of a writing instruction and a document\footnote{The writing instruction provides directions on what the model should do; the document provides context that the model needs to generate a response.}. Previous studies~\cite{hou2023ciphergpt,zhou2023textobfuscator,du2023dp} failed to protect the privacy within the {document} during the inference process in practical text generation tasks}. As shown in \autoref{tab:comparison}, 
CipherGPT~\cite{hou2023ciphergpt} utilized homomorphic encryption techniques in transformer-architecture models to enable inference on encrypted data. While these techniques can be used theoretically for privacy-preserving text generation tasks, they have limitations in practical applications due to the significant computational time and communication costs. TextObfuscator~\cite{zhou2023textobfuscator} and DP-Forward~\cite{du2023dp} added noise during data transmission in \textit{split learning}. However, they are mainly designed for classification tasks. Furthermore, they are unsuitable for black-box scenarios where the model owners, such as OpenAI~\cite{gpt4}, do not disclose details about the architectures of LLMs considering the intellectual property and commercial value of the models.

On the other hand, SANTEXT+~\cite{yue-etal-2021-differential} and CUSTEXT+~\cite{chen2023customized} leveraged \textit{local differential privacy} (LDP) techniques~\cite{dwork2006differential} to verbatim replace sensitive tokens in the text with semantically close tokens from a fixed token set, which is termed as \textit{static adjacency list} in the LDP context.
These methods are also designed for privacy-preserving classification tasks, which can tolerate considerable information distortion introduced by LDP noise. 
For the privacy-preserving text generation tasks~\cite{zhu2018texygen} investigated in this paper, even a slight information distortion in the \textit{prompt} can lead to incoherence and inconsistency in generated text, rendering SANTEXT+ and CUSTEXT+ not directly effective for such tasks. Additionally, the size of \textit{static adjacency list} equals the entire vocabulary in SANTEXT+, which is excessively large and increases the probability that perturbed tokens are semantically irrelevant to raw ones.
Moreover, our experimental results in~\autoref{fig:KNN250} demonstrate that CUSTEXT+ is vulnerable to embedding inversion attacks~\cite{qu2021natural}: {even in the extreme case where privacy parameter \(\epsilon\) is set to 0.01, an adversary can still recover 40\% of raw tokens in CUSTEXT+.} 
The rationale behind this phenomenon is that each token has a small \textit{static adjacency list} (which includes the raw token itself) in CUSTEXT+, thereby increasing the probability that the raw token will not be replaced.

\begin{table}[t]
\centering
\caption{An illustration of perturbed prompts and perturbed generation from GPT-4 by RANTEXT. The green text appears in both perturbed and original generations.} 
\label{tbl:PII_leak}
\resizebox{\columnwidth}{!}{
\begin{tabular}{l|l|l}
\toprule
\textbf{Method} &\textbf{Perturbed Prompt} & \textbf{Generated Text}\\
\midrule
Original & Sam lives in downtown Boston &and he enjoys walking through the historic streets.\\
\cmidrule{1-3}
$\epsilon=2.0$ & Lin living at city Barcelona &\bitone{and} she loves exploring local cafes and beaches.\\
$\epsilon=6.0$ & Mary lived in town Broadway &\bitone{and} she \bitone{enjoys} watching shows at \bitone{the} theater.\\
$\epsilon=10.0$ & Ben lives at urban Boston &\bitone{and he enjoys} exploring \bitone{the historic} landmarks.\\
\bottomrule
\end{tabular}
}
\end{table}

\mypara{Our Proposal}
%\zzk{Since we already introduced the LDP-based method and its shortcomings, here, we should directly describe how we address these shortcomings in our proposal.}
%\zzk{The above comment is not addressed, there is still a logic gap here. Also, the following paragraph is too long, try to separate it.}
To protect the privacy of the entire document during the inference process with black-box LLMs and address the information bias caused by LDP, we introduce a framework, \Name, for text generation tasks. The general concept of \Name is inspired by knowledge distillation~\cite{gou2021knowledge} and our \hyperlink{obs:example}{observation}. As illustrated in \autoref{tbl:PII_leak}, our \hyperlink{obs:example}{observation} is as follows: (1) The generation of the perturbed \textit{prompt} by LDP shares the same tokens across multiple parts of the generated text from the raw \textit{prompt}. (2) Furthermore, the number of shared tokens between them positively correlates with the privacy parameter $\epsilon$. This suggests that the perturbed generation result could potentially serve as a reference for a smaller language model to complete the generation task, distilling LLMs' generative capability. Based on the observation, \Name comprises a perturbation module and an extraction module. In the perturbation module, \Name employs a token-level LDP mechanism, such as SANTEXT+ and CUSTEXT+, to replace each token in the document with a new token, producing a perturbed \textit{prompt} shown in \autoref{fig:problem}. It uploads the perturbed \textit{prompt} to remote LLMs and obtains the perturbed generation result. In the extraction module, \Name deploys a local model that is lightweight and less capable than remote LLMs. This model extracts tokens from the perturbed generation result and reconstructs them into a generated text aligned with raw \textit{prompt}. Ultimately, \Name not only preserves the privacy of the \textit{prompt} but also leverages the capabilities of remote LLMs to enhance the local model’s output quality and complete the generation task.
%\zzk{We should briefly describe Assumption~\autoref{asump1} here, instead of referring the readers to other places.} %\zzk{Use present tense throughout the paper when describing our paper.} 
%\zzk{Use autoref instead of ref for all sections, figures, tables, and algorithms.}. , drawing ideas of knowledge distillation~\cite{gou2021knowledge} and retrieval-augmented generation~\cite{lewis2020retrieval}

%To address the vulnerability of SANTEXT+ and CUSTEXT+ in resisting the embedding inversion attack
To achieve a better balance between utility and privacy protection, we develop RANTEXT. It is a novel differentially private mechanism integrated into the text perturbation of \Name. RANTEXT introduces the concept of \textit{random adjacency list} for token-level perturbation. For each token, it employs the Laplace distribution~\cite{kotz2001laplace} to dynamically determine the size of the \textit{random adjacency list}, and then samples a new token from this list to replace raw tokens in the document. This approach enables RANTEXT to achieve a better trade-off between utility and privacy protection than existing methods do: (1) the \textit{random adjacency list} in RANTEXT is typically smaller than SANTEXT+'s \textit{static adjacency list}, which enhances the semantic utility of the perturbed text; (2) Compared with that in CUSTEXT+, the size of the \textit{adjacency list} in RANTEXT is generally larger, making it more difficult for an adversary to reconstruct the raw tokens. 

We conduct experiments on GPT-4~\cite{gpt4} for the evaluation of practical open-ended text generation tasks across three datasets. 
We found that existing attack strategies for differential privacy were not effective enough against RANTEXT. We propose an adaptive attack, the GPT inference attack, which leverages the capabilities of GPT-4 to reconstruct raw tokens. 

\mypara{Our\;Contributions}\!We\;summarize\;our main contributions:
\begin{itemize}
    \item We propose \Name, the first practical framework for privacy-preserving inference of black-box large language models, implementing differential privacy in text generation. 
    \item We develop RANTEXT, a novel exponential mechanism of local differential privacy integrated into document perturbation of \Name. It achieves a better balance between utility and privacy protection compared to existing baselines.
    \item We conduct experiments on three datasets tailored to practical open-ended text generation tasks in~\autoref{sec:utility}. Experimental results demonstrate that with \(\epsilon\) set to 3.0 and a 3.89GB local model, \Name achieves generation quality comparable to GPT-4 in terms of three metrics.
    \item We evaluate four classes of privacy threats in~\autoref{sec:privacy}. In particular, when we set the privacy parameter \( \epsilon \) to 6.0 and select the top 10 candidates for embedding inversion attack, RANTEXT offers an average privacy protection level exceeding 0.90, which is 3.35$\times$ higher that of CUSTEXT+ and 0.58$\times$ higher than that of SANTEXT+.
\end{itemize}

\section{Preliminaries}
\label{sec:preliminaries}
\subsection{Large Language Models}
\textit{Large language models} (LLMs) are advanced artificial intelligence systems trained on extensive datasets. 
They are designed to understand, generate, and interpret human language, demonstrating incredible versatility for various language-related tasks. Generally, LLMs generate text $Gen$ based on the prompt $Pro$ uploaded by the users. They come in different types, including closed-source commercial services like ChatGPT~\cite{ChatGPT} and Claude~\cite{claudeh}, as well as open-source models like Llama~\cite{touvron2023llama} and Vicuna~\cite{chiang2023vicuna}. In this paper, we focus on the closed-source LLMs and aim to address their privacy issues during the black-box inference in the open-ended text generation tasks.
 
Specifically, in the open-ended text generation task~\cite{welleck2019neural}, the role of these black-box LLMs is to continue generating text $Gen$ in accordance with the prompt $Pro$ for higher text generation quality based on multi-dimensional metrics. In detail, given a prompt $Pro = Ins \parallel Doc$ consisting of $Ins$ (\,fundamental writing instructions\,) and $Doc=\langle x_i \rangle_{i=1}^L $ (\,raw document composed of a sequence of $L$ tokens $x_i$, belonging to token vocabulary $V$), the LLMs commit to providing inference function $\text{Infer}(\cdot)\!:\!Pro\rightarrow Gen$ to generate text.
%\zzk{This subsection is too short, we can add more details, including different types of off-the-shelf LLMs, such as close models ChatGPT, Claude, Bard, and open-sourced models LLAMA, etc.}

\begin{table}[t]
\centering
\scriptsize
\caption{Notations and definitions. We indicate which elements are known to the adversary during the inference.}
\scalebox{1.2}{
\begin{tabular}{ l|| l }
\toprule
% {\bf Symbol} & {\bf Description}  \\ 
% \midrule
$U\!sr$ & User of LLMs \\
$Adv$ & LLMs as an adversary  \\
\midrule
$Ins$ & Instruction for text writing \\
$Doc$ & Raw document of the user  \\
$Doc_{p}$& Perturbed document of the user  \\
$Pro$ & Raw prompt of the user  \\
$Pro_p$ & Perturbed prompt of the user  \\
$Gen$ & Generation result of the raw prompt  \\
$Gen_p$&  Generation result of the perturbed prompt  \\
\midrule
$V$&  Token vocabulary  \\
$C\!_r$ & Random adjacency list of token  \\
$C\!_e$ & Random adjacent embeddings of token  \\
$C\!_s$ & Static adjacency list of token \\
$u$ & Scoring function of the exponential mechanism  \\
$M$ & Random mechanism of differential privacy \\
$\text{Infer}$ & Inference function for text generation of LLMs  \\
\bottomrule
\end{tabular}
}
\vspace{0cm}
\label{tab:notations}
\vspace{-0.3cm}
\end{table}

\subsection{(Local) Differential Privacy\& Exponential Mechanism}
% We briefly revisit two important definitions of differential privacy: $\epsilon$-Differential Privacy (\,$\epsilon$-DP\,), Exponential Mechanism (\,EM\,).

%\zzk{If I understand correctly, the technique in our paper is LDP instead of DP?
%Also, we need first describe the general idea of LDP before formally define it.}

Differential privacy~\cite{dwork2006differential} is a privacy protection concept. As one of its most popular models, \textit{$\epsilon$-local differential privacy} ($\epsilon$-LDP) allows data owners to locally perturb their data \cite{sharma2022finred} using the randomized mechanism $M(\cdot)$ before uploading it to any untrusted aggregator.

\begin{definition}[\textbf{$\epsilon$-Local Differential Privacy~\cite{kasiviswanathan2011can}}]
In $\epsilon$-LDP, given a privacy parameter $\epsilon \geq 0$, a randomized mechanism $M$ is $\epsilon$-LDP compliant if it satisfies the following condition for any two inputs $x, x' \in X$ and any possible output $y \in Y$:
\begin{equation}\label{dpbound}
\frac{\Pr[M(x) = y]}{\Pr[M(x') = y]} \leq e^\epsilon .
\end{equation}
\end{definition}

Typically, a smaller value of $\epsilon$ provides higher privacy protection at the cost of reduced data utility. Moreover, a critical definition here is the input set $X$. In previous NLP research~\cite{yue-etal-2021-differential,chen2023customized}, most researchers have posited that any pair of tokens in the vocabulary share the same input set $X$ and output set $Y$. We observe that such a definition leads to a challenge in the trade-off between utility and privacy. In this paper, we use \textit{random adjacency list} to redefine the input set of $\epsilon$-LDP in~\autoref{sec:random adjacency}.

\begin{definition}[\textbf{Exponential Mechanism~\cite{4389483}}]
For a given scoring function $u: X \times Y \rightarrow \mathbb{R}$, a randomized mechanism $M(\cdot)$ is $\epsilon$-LDP compliant if it satisfies the following condition for any input $x \in X$ and any possible output $y \in Y$:
\begin{equation}
\Pr[y|x] \propto \exp\left(\frac{\epsilon \cdot u(x,y)}{2\Delta u}\right),
\end{equation}
where the sensitivity $\Delta u$ is defined as:
\begin{equation}
\Delta u = \max_{x, x' \in X, y \in Y} |u(x,y) - u(x',y)|.
\end{equation}
\end{definition}
The scoring function $u$ is various in different scenarios. Typically, we can adjust the upper bound of $u$ to set $\Delta u $ to a specific real number, where $\Delta u $ represents the sensitivity of the scoring function $u$. Similarly, the smaller the value of $ \epsilon $, the higher the security of privacy protection capability, but the lower the utility of the data. When a smaller $ \epsilon $ is chosen, the scoring function $u(x,y) $ no longer plays a decisive role in the output probability of any perturbation result. 

%\zzk{There should be some sentences to connect the following contents.}
%The above are two key definitions of differential privacy. To formally describe the appropriate type of differential privacy algorithms for text perturbation, we propose the following definition:
%\begin{definition}[\textbf{$\epsilon$-Document Privacy}]
%Given a privacy parameter $\epsilon \geq 0$, a randomized mechanism $M(\cdot)$ of differential privacy is $\epsilon$-document privacy if it holds that:
%for any word $x$ in a vocabulary $V$ or a token $x$ in the token vocabulary $V_t$, $y$ and $z$ belonging to the static adjacency list of $x$ or random adjacency list of $x$, we have
%\begin{equation}
%d(x, y) \geq d(x, z) \Rightarrow \Pr[M(x)=y] \leq \Pr[M(x)=z],
%\end{equation}
%where $d(\cdot)$ measures the semantic similarity between two inputs, with a smaller output indicating greater similarity. 
%\end{definition}

%The definition of $\epsilon$-Document Privacy suggests that differential privacy in text perturbation should guarantee a higher probability of outputting semantically closer words. The perturbation in \Name is based on $\epsilon$-differential privacy, thereby ensuring the semantic utility of perturbed text by 
%sampling process.

%\subsection{Privacy-preserving LLMs}
%\zzk{We should use a separate section before the conclusion section to discuss the related work in more detail.}

\section{Problem Statement}

% \zzk{Put together the adversary's goal}

\subsection{Threat Model}
\label{sec:threat-model}
We consider the scenario where the LLM platform, such as ChatGPT, is an honest but curious adversary, referred to as $Adv$. 
A user, denoted as $U\!sr$, intends to upload a prompt and invoke the inference service $\text{Infer}(\cdot)\!:\!Pro\rightarrow Gen$ of $Adv$ to complete the text generation tasks, which are poorly executed by open-source models. 
Here, $Gen$ denotes the text generated by $Adv$. The uploaded prompt, $Pro = Ins \parallel Doc$ represents the raw prompt of $U\!sr$ consisting of $Ins$ (\,fundamental writing instructions\,) and $Doc=\langle x_i \rangle_{i=1}^L $ (\,raw document composed of a sequence of $L$ tokens $x_i$, belonging to token vocabulary $V$\,). 
\begin{figure*}[t]
\centering
\includegraphics[width=\textwidth]{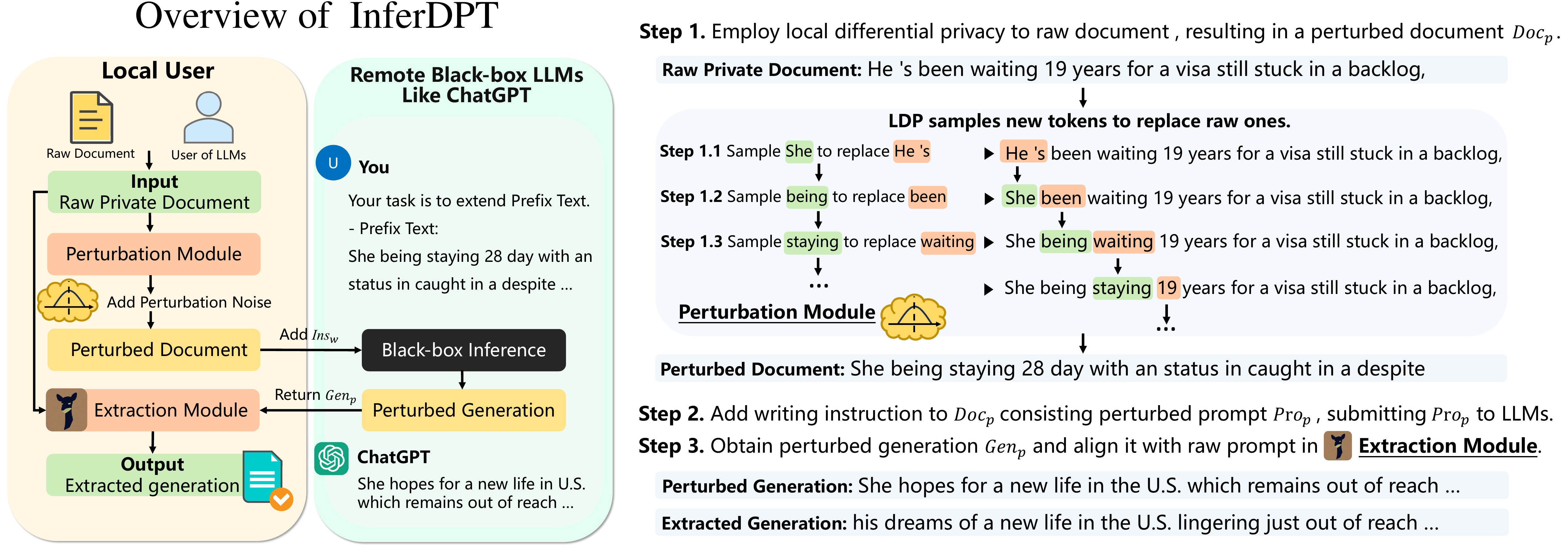}
\caption{The overview of \Name. It consists of (1) a perturbation module that samples new tokens to replace the raw ones in $Doc$ via LDP and (2) an extraction module that locally aligns the perturbed generation with the raw document.}
\label{fig:InferDPT}
\end{figure*}
\begin{figure}[t]
\centering
\includegraphics[width=0.95\columnwidth]{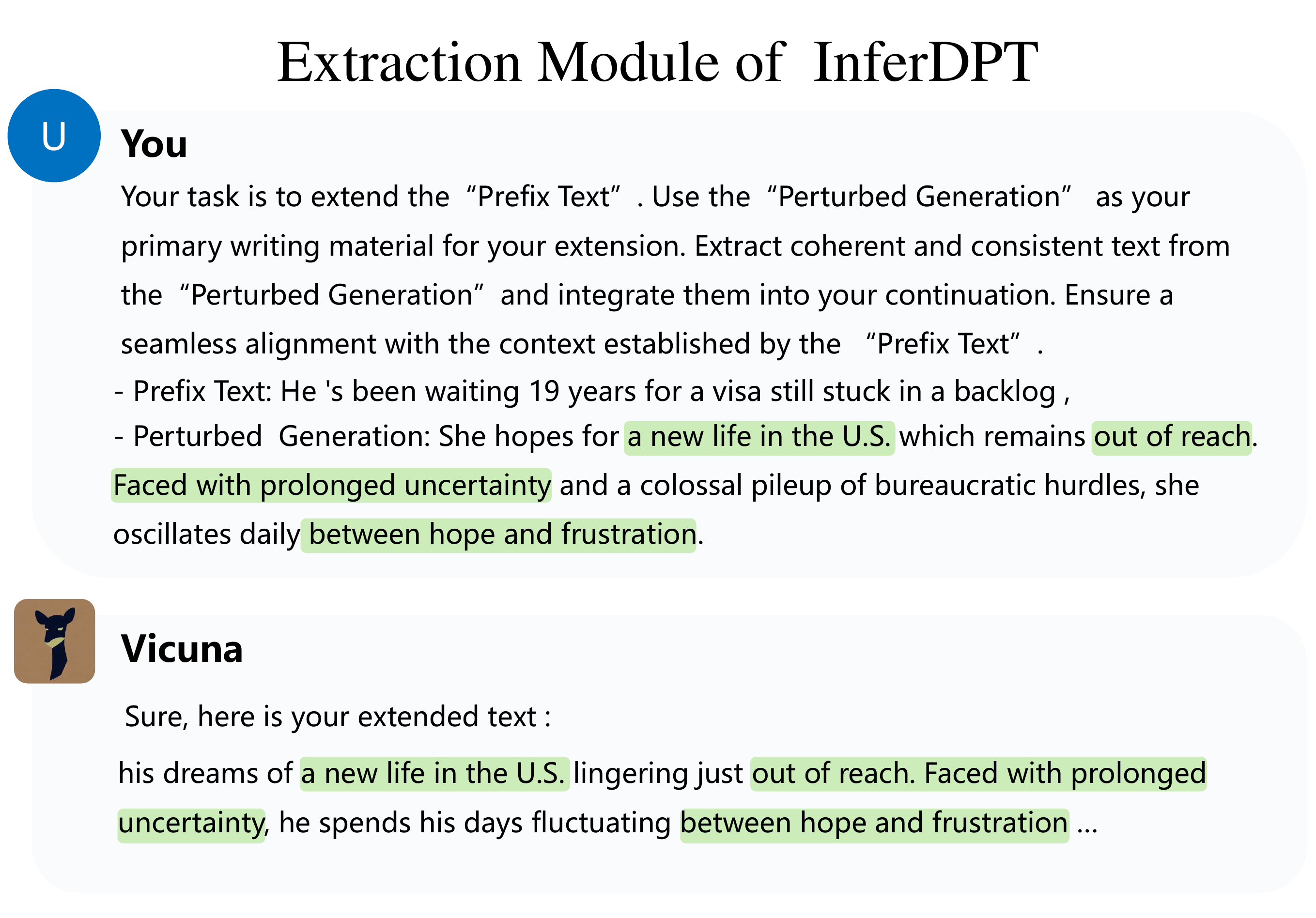}
\caption{The extraction module employs a smaller language model locally to extract text from the perturbed generation. It then reconstructs this text into an output that aligns with the raw document. We mark the text in green to indicate that it is identical in both the perturbed and extracted generations.}
\label{fig:extraction}
\end{figure}

Following previous works~\cite{zhou2023textobfuscator,qu2021natural}, the privacy information probably pertains to each token. \ul{To protect each piece of the token in the raw document $Doc$}, $U\!sr$ employs differential privacy~\cite{dwork2006differential} to $Doc$, resulting in a perturbed document $Doc_{p}$. Consequently, $U\!sr$ uploads the perturbed prompt $Pro_p = Ins \parallel Doc_{p}$. Furthermore, $U\!sr$ can deploy a less capable language model than LLMs. To preserve the model's commercial value, $Adv$ does not reveal the internal architecture or parameters of the LLMs, but only exposes its token vocabulary $V$ to $U\!sr$ for the purpose of billing verification during the inference process.
    
The goal of $Adv$ is to reconstruct every piece of the token in the raw document $Doc$ from $Doc_{p}$. $Adv$ is expected to launch attacks using vulnerabilities in LDP, aiming to recover each token in the document $Doc$, based on the perturbed version $Doc_{p}$. Additionally, we assume that $Adv$ is fully informed about the details of the differential privacy algorithm.

%Taking a prompt for drafting an article as an example, $Adv$ is committed to executing the text generation tasks but may intend to steal its experimental results for unauthorized collection. 

%Additionally, we assume that $Adv$ is fully informed about the details of the differential privacy algorithm, except for the raw document $Doc$. Given a perturbed prompt $Pro_p = Ins \parallel Doc_{p}$ uploaded by $U\!sr$, the goal of $Adv$ is to reconstruct every piece of the token in the raw document $Doc$ from $Doc_{p}$. $Adv$ is expected to launch attacks using vulnerabilities in LDP, aiming to recover each token in the document $Doc$, based on the perturbed version $Doc_{p}$. 
%This assumption underscores a stringent requirement for differential privacy algorithms. Any flaw in the design of these algorithms could enable $Adv$ to recover a substantial amount of the original information by exploiting DP vulnerabilities.
\autoref{tab:notations} summarizes notations frequently used in this paper.

\subsection{Existing Solutions and Limitations}\label{sec:esw}
%\zzk{Discuss SANTEXT+ and CUSTEXT+ in detail and what are their shortcomings.}

Existing solutions, such as SANTEXT+~\cite{yue-etal-2021-differential} and CUSTEXT+~\cite{chen2023customized}, focus on privacy-preserving model training in classification tasks:
\begin{itemize}
\item {SANTEXT+} implements local differential privacy (LDP)~\cite{alvim2018local} during the training in classification tasks. It substitutes the raw tokens with newly sampled ones from a \textit{static adjacency list}. However, this \textit{static adjacency list} in SANTEXT+ equals the entire token vocabulary and is excessively large. Consequently, there exists a high probability that the perturbed token may be semantically irrelevant to the raw one, leading to diminished utility of the perturbed text.
\item {CUSTEXT+} perturbs each token, excluding stopwords~\cite{stopwords}, during training in classification tasks. Compared to SANTEXT+, it reduces the size of \textit{static adjacency list} to a small number (default 20) for better utility of LDP. However, this small \textit{static adjacency list} increases the probability that raw tokens will not be replaced, resulting in privacy leakage.
\end{itemize}

To protect the privacy of documents during inference in text generation tasks and address the information distortion introduced by LDP noise, we introduce a framework, \Name, (\autoref{sec:infer}). We also propose an exponential mechanism, RANTEXT (see \autoref{sec:rantext}), which offers a better trade-off between utility and privacy protection compared to existing SANTEXT+ and CUSTEXT+. 
%as a solution to address the vulnerabilities of SANTEXT+ and CUSTEXT in resisting the embedding inversion attack. 

%\vspace{-1mm}

\section{The \Name Framework}\label{sec:infer}
\subsection{Overview}

%\zzk{The current logic is still a bit strange, there is no connection on how do we address the drawback of SANTEXT+ and CUSTEXT+.}
We introduce \textsf{\Name}, a framework designed for privacy-preserving LLMs inference in text generation tasks. As shown in ~\autoref{fig:InferDPT}, \Name is consisting of two modules:
%\zzk{Provide more details for each of the modules.}
\begin{itemize}
\item \textbf{Perturbation Module:~protecting privacy}. It generates a perturbed document by replacing each token in $Doc$ with one close to embedding distance and sampled by LDP.
\item \textbf{Extraction Module:~maintaining utility}. It extracts coherent and consistent text from perturbed generation and reconstructs them into an output aligned with the raw prompt by a local language model, less capable than black-box LLMs.
\end{itemize}

The design of \Name faces two main challenges in black-box inference. (1) \textbf{Providing strong privacy protection for the raw document $Doc$}. To solve this privacy challenge, the perturbation module of \Name utilizes a differentially private mechanism to sequentially replace each token in the raw document $Doc$ with alternatives close in embedding distance.
(2) \textbf{Maintaining the utility of the text under semantic perturbation}. This is more tough than the first one. To solve this challenge, we conducted abundant experiments about the generation of the perturbed document using LDP on LLMs. Specifically, we perturb each token in the document with a newly sampled one close in the embedding distance by LDP, resulting in a perturbed document. We discovered that the generation of this perturbed document includes numerous tokens found in the generation of the raw document. For example, we collect the tokens appearing in the generation of the raw document $Doc$ in \autoref{fig:InferDPT}, termed a set $\{hopes,\,new,\,dreams,\cdots\}$. We find that the generation of the perturbed document $Doc_p$ (depicted in \autoref{fig:InferDPT}) contains tokens within the set $\{hopes,\,new,\,dreams,\cdots\}$. Moreover, this overlap increases gradually as the perturbation decreases.

\begin{figure*}[t]
\centering
\begin{minipage}{1.0\textwidth}
  \centering
  \includegraphics[width=\linewidth]{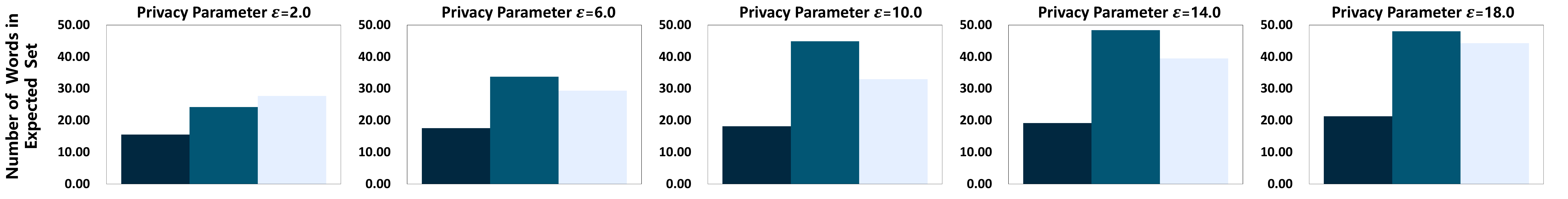}
\end{minipage}%
\vspace{5pt} % 调整这里的间距
\begin{minipage}{\textwidth}
  \centering
  \includegraphics[width=0.9\linewidth]{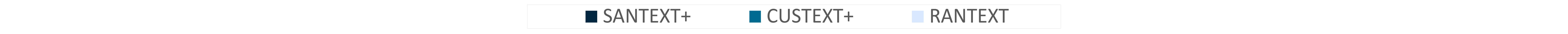}
\end{minipage}
\caption{The number of tokens from the non-private and private generation of GPT4 using three mechanisms that belong to the \textit{Expected set}.}\label{fig:vali_ass}
\vspace{-10pt}
\end{figure*}

To formally describe this phenomenon, we propose the following Observation about the perturbed output by LDP.
%Formally, we make the following observation that conjectures the outcome of the inference on the perturbed prompt $Pro_p$.

%\zzk{The following definitions and assumptions should be put in the perturbation module part.}
\subsection{Key Observations}\label{sec:assumption}

%\zzk{Are there observation 2? If no, we should not number it, just observation is sufficient.}
\mypara{Observation}\hypertarget{obs:example}{}
Let $V$ be a token vocabulary. We define $d(\cdot)$ as a function that quantifies the semantic similarity between two tokens, where smaller output values indicate greater similarity. Let $ M(\cdot)$ denote a randomized function of LDP that satisfies: 
\begin{equation}
d(x, y) \geq d(x, z) \Rightarrow \Pr[M(x)=y] \leq \Pr[M(x)=z],
\end{equation}
\textit{where tokens $x,y,z\in V$}.

Let tokens $x_i,y_i \in V$. $Doc_{p}=\langle  y_i \rangle_{i=1}^{L}$ denotes the perturbed document of the raw document $Doc=\langle  x_i \rangle_{i=1}^{L}$ by $y_i=M(x_i)$. The expression $\text{Infer}(Ins \parallel Doc)=\langle  h_i^{(j)}\rangle_{i=1}^{K}$ 
represents the generation result of the $j$-th inference on $Doc$, consisting of tokens $h_i^{(j)}\in V$. Given $Doc_{p}$, the perturbed generation $Gen_p =\langle g_i \rangle_{i=1}^{K} =\text{Infer}(Ins \parallel Doc_p)$ consists of tokens $g_i \in V$ and satisfies the following condtion:
\begin{equation}
    \textit{Expected~set} = \bigcup_{j=1}^{N} \{ h_i^{(j)} |\, h_i^{(j)} \in \text{Infer}(Ins \parallel Doc) \},
\end{equation}
\vspace{-14pt}
\begin{align}
    \hspace{-4pt}\textit{Intersection} = \{g_i|g_i\! \in\! \textit{Expected\;set} \, \text{and} \, g_i \notin \textit{stopwords}\text{\footnotemark}\},
\end{align}
\footnotetext{Stopwords~\cite{stopwords} are common words usually ignored in text analysis due to their limited informational values.}
\vspace{-8pt}
\begin{equation}
   \text{Corr}(\text{Count}(\textit{Intersection}),\,\epsilon) > 0,
\end{equation}
\textit{where $N$ is a positive integer; the function }\text{Count}\textit{$(\cdot)$ counts the size of a set; the function }\text{Corr}\textit{$(\cdot)$ measures correlation coefficient between two variables, with values ranging from -1 (negative correlation) to 1 (positive correlation).}

\mypara{Implication}This \hyperlink{obs:example}{Observation} states that if the \textit{Expected set} is constructed from tokens in the results of $N$ iterations of raw prompt, then the presence of tokens from the perturbed generation within the \textit{Expected set} will positively correlate with $\epsilon$. This implies that smaller perturbations to $Doc$ lead to higher consistency between the perturbed generation and the raw generation. To verify this \hyperlink{obs:example}{Observation}, we carried out the following experiments with GPT-4~\cite{gpt4}. %This implies that when using the 

\mypara{Empirical Validation}We got the $Expected~set$ by collecting 100 tokens from the output of the raw prompt with GPT-4 generated 100 times on the CNN/Daily Mail dataset~\cite{hermann2015teaching}. The raw prompt consists of a fundamental writing instruction and a raw document of 50 tokens shown in ~\autoref{fig:InferDPT}. We utilized SANTEXT+~\cite{yue-etal-2021-differential}, CUSTEXT+~\cite{chen2023customized}, and RANTEXT introduced in~\autoref{sec:rantext} to generate perturbed outputs of 100 tokens from GPT-4 under various values of $\epsilon$. We counted the number of tokens from the perturbed and non-private generation of GPT-4 that belong to the $Expected~set$.

\autoref{fig:vali_ass} shows the experimental results. We can see that with the increase of $\epsilon$ and reduction of perturbation, the number of tokens in the $Expected~set$ of the three mechanisms has increased. This validates \hyperlink{obs:example}{observation}, confirming that the number of tokens from the $Expected~set$ appearing in the perturbed generation positively correlates with $\epsilon$.

In conclusion, our observation suggests that the perturbed generation result shares the same tokens in multiple parts of the non-perturbed generation result. However, the perturbed generation result lacks some information in the raw document, which LDP replaces. To address this, \Name employs an extraction module to extract related text from the perturbed generation result as an output reference, distilling the generation capabilities of LLMs. It then reconstructs them into a generated text aligned with the raw document. In the following subsections, we will delve into the perturbation module and the extraction module of \Name.

\subsection{Perturbation Module}
\floatname{algorithm}{Algorithm}
\begin{algorithm}[t]
\small
\caption{Perturbation Module}
\label{alg:perturbation_module}
\begin{algorithmic}[1]
\REQUIRE Document $Doc=\langle x_i \rangle_{i=1}^L$,  random mechanism $ M(\cdot) $, static adjacency list $C\!_s(\cdot)$, \textit{random adjacency list} $C\!_r(\cdot)$;
\ENSURE Perturbed document $Doc_p$;
\STATE Initialize $Doc_p \gets \emptyset$;
\FOR{$i = 1 \ to \ L$}
    \STATE Compute $C\!_s(x_i)$ or $C\!_r(x_i)$; %\Comment{Define Adjacent word}
    \STATE Sample $y_i \sim M(x_i,\,C\!_s(x_i))$ or $y_i \sim M(x_i,\,C\!_r(x_i))$; %\Comment{Sample from $C_s(x_i)$}
    \STATE Append $y_i$ to $Doc_p$; %\Comment{Replace $x_i$ with $x_i'$}
\ENDFOR
\STATE Output $Doc_p$;
\end{algorithmic}
\end{algorithm}
After perturbation, $U\!sr$ uploads a perturbed prompt $Pro_p = Ins \parallel Doc_p$ (consisting of a writing instruction $Ins$ and a perturbed document $Doc_p$) to remote LLMs. The LLMs then return perturbed generation $Gen_p =\text{Infer}(Pro_p)$ to $U\!sr$. 
\begin{figure*}[h]
\centering
%\hspace{-0.2cm}
\includegraphics[width=\textwidth]{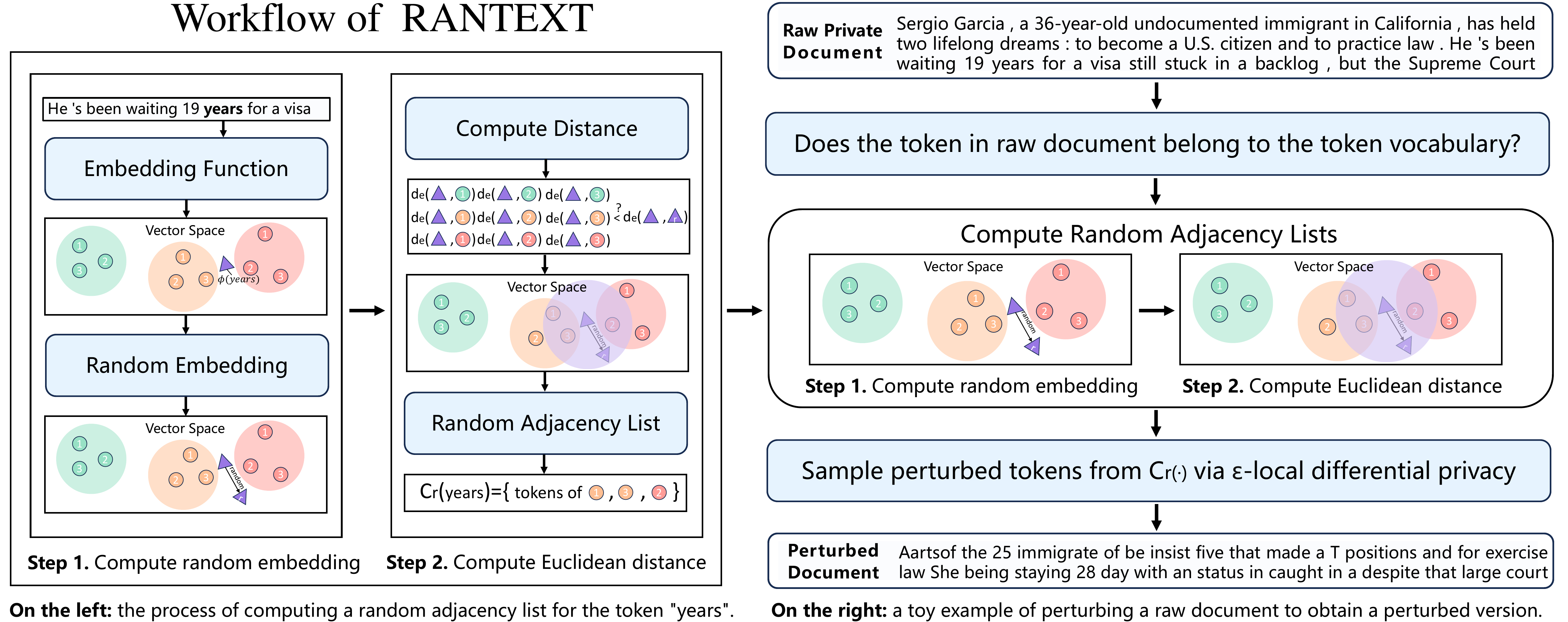}
\caption{The workflow of RANTEXT. It comprises two steps: (1) computing random adjacency lists and (2) sampling perturbed tokens via \(\epsilon\)-LDP.}
\vspace{-10pt}
\label{fig:RANTEXT}
\end{figure*}
The perturbation module of \Name generates a perturbed document $Doc_p$ from an input document $Doc$. Specifically, it replaces each token in $Doc$ with new tokens sampled by the randomized mechanism $ M(\cdot) $ of LDP from a \textit{token set}. In previous works \cite{yue-etal-2021-differential,chen2023customized}, this \textit{token set} is static (termed the \textit{static adjacency list} $C\!_s$), whereas in our proposed RANTEXT, it is random (termed the \textit{random adjacency list} $C\!_r$). Typically, $C\!_s$ (or $C\!_r$) consists of tokens that are close to the embedding of the token $ x_i \in V $ to be replaced. Given a document $ Doc = \langle x_i \rangle_{i=1}^L $ composed of $L$ tokens $ x_i \in V $, the perturbation module replaces each $ x_i $ with a random output $ y_i = M(x_i,\,C\!_s(x_i)) $ or $ y_i = M(x_i,\,C\!_r(x_i)) $, resulting in a perturbed document $ Doc_p = \langle y_i \rangle_{i=1}^L$ (as shown in~\autoref{fig:InferDPT}). The detailed process of the perturbation module is outlined in \autoref{alg:perturbation_module}. In the implementations, \Name adopts three mechanisms of LDP: SANTEXT+ \cite{yue-etal-2021-differential}, CUSTEXT+ \cite{chen2023customized}, and RANTEXT (detailed in the following \autoref{sec:rantext}).

While LDP perturbs sensitive text, it is important to note that an excessively large $\epsilon$ in LDP increases the risk of privacy leakage from the perturbed text. This is because LDP perturbs raw tokens to more semantically close tokens as $\epsilon$ increases.
We experimentally demonstrate this risk by calculating the synonymous token proportion and the embedding distance (between the raw tokens and their perturbed tokens) with various $\epsilon$ values. As shown in  \autoref{tbl:synonym} and \autoref{tbl:cos_emb}, as $\epsilon$ increases, the synonym proportion grows and the Euclidean distance decreases, indicating greater semantic similarity between the raw tokens and the perturbed tokens.
\subsection{Extraction Module}
As previously mentioned, the perturbation module disturbs each token and key information in the raw document $Doc$, making it difficult for an adversary to reconstruct the raw tokens from $Doc_p$ or $Gen_p$. However, this perturbation also leads to inconsistency and partial incoherence of semantics between $Gen_p$ and $Doc$, as illustrated in \autoref{fig:InferDPT}.

To obtain the aligned generation of the raw document $Doc$, the extraction module of \Name deploys a local language model that is considered trustworthy and does not pose any privacy leakage issues. This local model is smaller and less powerful than remote black-box LLMs, facilitating easier implementation under limited resources. As shown in \autoref{fig:extraction}, $U\!sr$ inputs the raw document $Doc$ and the perturbed generation $Gen_p$ into this local model. This model is tasked with extracting coherent and consistent text from $Gen_p$ and integrating it into the continuation of $Doc$, ensuring an aligned output. Although the local model can generate aligned content independently, the generation quality is not satisfactory due to its limited capabilities. However, with the perturbed generation $Gen_p$, the local model distills the capacity of the remote black-box LLMs. The details of the prompt utilized in the extraction module can be found in \hyperref[app:B]{Appendix A}. 

\begin{table}[t]
\centering
\caption{Proportion of synonyms between the raw tokens and their perturbed versions~\cite{stopwords}.}
\label{tbl:synonym}
\vspace{-2pt}
 \resizebox{0.9\columnwidth}{!}{
\begin{tabular}{l||cccc}
\toprule
 \multirow{2}{*}{\textbf{Method}} 
& \multicolumn{4}{c}{\textbf{Synonym Proportion$\downarrow$}} \\ \cline{2-5} 
 & 
\rule{0pt}{2.5ex} $\epsilon=2.0$ & $\epsilon=6.0$  & $\epsilon=10.0$ &$\epsilon=14.0$\\
 \midrule
SANTEXT$^+$
& $0.371$ & $0.373$ & $0.374$ &$\mathbf{0.375}$\\ 
 CUSTEXT$^+$
& $0.441$ & $0.697$ & $0.907$ &$0.985$\\ 
 RANTEXT
& $\mathbf{0.013}$ & $\mathbf{0.049}$ & $\mathbf{0.147}$ &$0.378$\\ 
 \bottomrule
\end{tabular}
}
\vspace{-6pt}
\end{table}
\vspace{0.1cm}
\begin{table}[t]
\centering
\caption{Euclidean distance between the embeddings of tokens and their perturbed versions.}
\label{tbl:cos_emb}
\vspace{-2pt}
 \resizebox{0.9\columnwidth}{!}{
\begin{tabular}{l||cccc}
\toprule
 \multirow{2}{*}{\textbf{Method}} 
& \multicolumn{4}{c}{\textbf{Euclidean Distance$\uparrow$}} \\ \cline{2-5} 
 & 
\rule{0pt}{2.5ex} $\epsilon=2.0$ & $\epsilon=6.0$  & $\epsilon=10.0$ &$\epsilon=14.0$\\
 \midrule
SANTEXT$^+$
& $3.081$ & $2.775$ & $2.756$ &${2.750}$\\ 
 CUSTEXT$^+$
& ${2.862}$ & ${1.732}$ & ${0.553}$ &${0.118}$\\ 
 RANTEXT
& $\mathbf{4.317}$ & $\mathbf{4.133}$ & $\mathbf{3.667}$ &$\mathbf{2.807}$\\ 
 \bottomrule
\end{tabular}
}
\vspace{-6pt}
\end{table}

Based on the above description, we have a panoramic view of \Name. It is noted that the perturbation module can adopt existing differentially private mechanisms such as SANTEXT+~\cite{yue-etal-2021-differential} and CUSTEXT+~\cite{chen2023customized}. However, these two have drawbacks either in terms of utility or security, as analyzed in~\autoref{sec:esw}. To address these problems, we introduce RANTEXT in the following section.

\section{The RANTEXT Mechanism}~\label{sec:rantext}
\vspace{-24pt}
\subsection{Overview}
We design RANTEXT to address the utility and vulnerability problems of previous differentially private mechanisms~\cite{yue-etal-2021-differential,chen2023customized}. As shown in~\autoref{fig:RANTEXT}, RANTEXT comprises two steps:
\begin{itemize}
    \item \textbf{Compute Random Adjacency Lists.} This step computes a \textit{random adjacency list} for each raw token via two operations: computing random embedding and Euclidean distance. Any tokens in \textit{random adjacency list} share the same input set.
    \item \textbf{Sample Perturbed Tokens via \(\epsilon\)-LDP.} This step samples a perturbed token for each raw token and replaces the raw token in the document from its \textit{random adjacency list} via \(\epsilon\)-LDP, obtaining the perturbed document.
\end{itemize}

As mentioned in Section \ref{sec:threat-model}, LLMs expose their token vocabulary $V$ for billing verification of inference service. 
Utilizing the token vocabulary $V$ and Byte Pair Encode (BPE) algorithm~\cite{sennrich2015neural}, users can obtain the $tokenizer(\cdot)$ of LLMs. 

Given a raw document $Doc$, RANTEXT first uses the $tokenizer(\cdot)$ algorithm of LLMs to turn the text of $Doc$ into tokens $\langle {x_i} \rangle_{i=1}^L$, where $x_i \in V$:
\begin{equation}
T\!okenset=\langle {x_i} \rangle_{i=1}^L =tokenizer(Doc). 
\end{equation}

To preserve the privacy of $Doc$, RANTEXT discards the tokens of $Doc$ that do not belong to $V$ and employs an exponential mechanism to subsequently replace each remaining token with one close in embedding distance from its exclusive \textit{random adjacency list}:
\begin{equation}
   {r_i} =M(x_i,C\!_r(x_i)),
\end{equation}
\begin{equation}
    T\!okenset_p=\langle {r_i} \rangle_{i=1}^l=\langle {M(x_i,C\!_r(x_i))} \rangle_{i=1}^l,
\end{equation}
\noindent where token $r_i \in V$, $T\!okenset_p$ represents the perturbed token set, and $C\!_r(x_i)$ represents the \textit{random adjacency list} of  $x_i$. RANTEXT concatenates the tokens in a perturbed token set $T\!okenset_p$ to obtain a perturbed document $Doc_p$, thereby providing privacy protection.

\subsection{Compute Random Adjacency Lists}\label{sec:random adjacency}

To formally define the \textit{random adjacency list}, we first give a definition of \textit{random adjacent embeddings}:

\begin{definition}[\textbf{Random Adjacent Embeddings}]

 Given token $t \in V$, its random adjacent embeddings are defined as follows:
\begin{equation}\label{eq15}
C\!_e(t) = \{ \text{eb} | d_e\left(\text{eb}, \phi(t)\right) < d_e(\hat{\phi}(t),\phi(t)),\text{eb}\in \mathbb{R}^N\},
\end{equation}
where $eb \in \mathbb{R}^N$ represents any $N$-dimensional vector within the real number domain. The function $d_e(\cdot)$ is utilized to compute the distance between two vectors and is defined as $d_e({a}, {b}) = \sqrt{\sum_{i=1}^n ({a}_i - {b}_i)^2}$. The function $\phi : V \rightarrow \mathbb{R}^N$ maps any given token to a vector in the $N$-dimensional real number vector space. The function $\hat{\phi}(t) = {\phi}(t) + Y$, where 
the random vector $Y$ satisfies the probability density:
\begin{equation}\label{eqdp}
Y \sim f(x)=  \frac{Z}{2\Delta{\phi}}\cdot\exp\left(-\frac{Z\cdot|x |}{\Delta{\phi}}\right),\\
\end{equation}
\begin{equation}
Z = 
\begin{cases}
  ~~{\epsilon} & \text{if}~\epsilon<2,\\
  {a \log(b \cdot \epsilon + c)+d} & \text{otherwise},
 \end{cases}\\
\end{equation}
$where\;\Delta{\phi}\;is\;the\;sensitivity\;of\;function\;\phi(\cdot),\;a\approx\;0.0165,\;b\;\approx 19.0648,\;c\;\approx-38.1294,\;d\approx9.3111$\,.
\end{definition}
%\frac{\log(\epsilon)}{10} + \frac{4}{3}

Given a token $t\in V$ to compute its random adjacent embeddings, we need to complete the two-step computation:\\
\textbf{\textit{Step 1. Compute the random embedding.}} 
We construct the random vector $Y$ utilizing the Laplace distribution~\cite{kotz2001laplace}. We add $Y$ independently to each dimension of $\phi(t)$, obtaining 
the random embedding $\hat{\phi}(t) = {\phi}(t) + Y$ of raw private token $t$. \\
\textbf{\textit{Step 2. Compute the Euclidean distance.}}
We compute~the Euclidean distance between ${\phi}(t)$ and $\hat{\phi}(t)$, referred to $d_e(\hat{\phi}(t),\phi(t))$. The random adjacent embeddings consist of those embeddings whose Euclidean distance to \(\phi(t)\) is shorter than \(d_e(\hat{\phi}(t), \phi(t))\). 

\begin{algorithm}[t]
\caption{RANTEXT Mechanism}
\label{alg:RANTEXT}
\begin{algorithmic}[1]
\REQUIRE Token set $T\!okenset=\langle {x_i} \rangle_{i=1}^L$, token vocabulary $V$, privacy parameter $\epsilon$, embedding function $\phi(\cdot)$, distance function $d_e(\cdot)$, random vector $Y$;
\ENSURE Perturbed document $Doc_p$;
\STATE Initialize $T\!okenset_p \gets \emptyset$;
\FOR{$i = 1$ to $L$}
    \IF{$x_i \notin V$}
        \STATE Discard the token $x_i$\;;
        \STATE Continue;
    \ENDIF
    \STATE Sample a random vector $Y$;
    \STATE Compute embedding $eb_t \gets \phi(x_i)$;
    \STATE Compute random embedding $eb_n \gets eb_t + Y$;
    \STATE Compute Euclidean distance $d_{threshold} \gets d_e(eb_n, eb_{t})$;\!
    \STATE $C\!_e({x_i}) = \{ {eb}\; |\; d_e\left({eb}, eb_{t}\right) < d_{{threshold}}\,, \;eb\in \mathbb{R}^N\}$;
    \STATE $C\!_r({x_i}) = \left\{{x_i}'| {\phi}({{x_i}}') \in C\!_e({x_i}),{x_i}' \in V \right\}$;
    \FOR{each ${x_i}' \in C\!_r(x_i)$}
        \STATE \(d_{{x_i}'} \leftarrow d_e(\phi(x_i), \phi({x_i}'))\);
        \STATE Scoring function \(u(x_i, {x_i}') \leftarrow 1 - d_{{x_i}'} / d_{threshold}\);
        \STATE $p_{total} \gets p_{total} + \exp\left( \epsilon / 2 \cdot u(x_i, {x_i}') \right)$;
    \ENDFOR
    \FOR{each ${x_i}'' \in C\!_r(x_i)$}
        \STATE $p({x_i}'' | x_i) \gets \exp\left( \epsilon / 2 \cdot u(x_i, {x_i}'') \right) / p_{total}$;
    \ENDFOR
    \STATE Sample from \textit{random adjacency list} $r_i \sim p({x_i}'' | x_i)$;\!\!
    \STATE Append new token $r_i$ to perturbed token set $T\!okenset_p$;
\ENDFOR
\STATE Concatenate $T\!okenset_p=\langle {r_i} \rangle_{i=1}^L$ obtaining $Doc_p$
\STATE Output perturbed document $Doc_p$;
\end{algorithmic}
\end{algorithm}

We use $Y$ to dynamically determine the size of the \textit{random adjacency list}. The detailed construction process of the random vector $Y$ can be found in \hyperref[sec:noise]{Appendix B}.

With the definition of random adjacent embeddings, we give the definition of the \textit{random adjacency list}:
\begin{definition}[\textbf{Random Adjacency List}]
Given a token $t \in V$, its \textit{random adjacency list} is defined as follows:
\begin{equation}\label{eq17}
C\!_r(t) = \left\{t'| {\phi}({t}') \in C\!_e(t),t' \in V\right\}.
\end{equation}
\end{definition}
% Algorithm code can stay as is since it's technical material

Given a token $t \in V$, its \textit{random adjacency list} is composed of any token $t'$ in the token vocabulary $V$, whose embedding $\phi(t')$ has a Euclidean distance to $\phi(t)$ shorter than the Euclidean distance between $t$'s random embedding and $t$'s embedding $\phi(t)$.

The design of the \textit{random adjacency list} in RANTEXT obeys the following theorem:
\begin{theorem}\label{theorem:1}
Given a token $t \in V$ and any token $t' \in V$, there exists a \textit{random adjacency list} $C\!_r(t)$ of RANTEXT satisfying $t' \!\in C\!_r(t)$.
\end{theorem}
\autoref{theorem:1} is proven in \hyperref[app:proof]{Appendix C}. It demonstrates that a token $t$ can be substituted with any token $t'\in V$ in RANTEXT, thereby increasing the difficulty for adversaries to reconstruct the raw tokens. Moreover, the \textit{random adjacency list} addresses the utility problem of the perturbed text in SANTEXT+. Although the theoretically maximum size of the \textit{random adjacency list} is equivalent to the size of $V$, it is typically smaller than that in terms of probability. This reduces the likelihood that the perturbed token is semantically irrelevant. 

Furthermore, experimental results in the following \autoref{sec:utility} demonstrate that the \textit{random adjacency list} in RANTEXT is generally larger than the \textit{static adjacency list} in CUSTEXT+, which solves the vulnerability of CUSTEXT+~\cite{chen2023customized} to the embedding inversion attack. 

\subsection{Sampling Perturbed Tokens via \(\epsilon\)-LDP} 
In SANTEXT+~\cite{yue-etal-2021-differential}, a proportion of tokens is not perturbed by LDP. To solve the privacy leakage issue in the raw text, RANTEXT perturbs every piece of the token in $T\!okenset=\langle {x_i} \rangle_{i=1}^L$. To perturb token $x_i$, RANTEXT employs the exponential mechanism~\cite{4389483}, which satisfies \(\epsilon\)-LDP, to select a new token from $C\!_r({x_i})$ to replace the original one. For any special token $t_s \notin V$, RANTEXT discards it, to ensure there is no special token leakage in $Doc_p$.

To guarantee the utility of the perturbed document, the random mechanism $ M(\cdot)$ of the exponential mechanism in RANTEXT is required to satisfy:
\begin{equation}
d(x, y) \geq d(x, z) \Rightarrow Pr[M(x)=y] \leq Pr[M(x)=z],
\end{equation}
\textit{where $x \in$ $V$, and $y$ and $z$ belong to the \textit{random adjacency list} of $x$. $d(\cdot)$ measures the semantic similarity between two inputs, with a smaller output indicating greater similarity.} 

To fulfill that, the scoring function $u(\cdot)$ of the random mechanism $ M(\cdot)$ in RANTEXT is described as follows:

\textit{Given a token $t$, RANTEXT considers that any two tokens in $C\!_r({t})$ share the same input set and output set during the perturbation of token $t$.}
\textit{Given any two tokens $x, y \in C\!_r(t)$, the scoring function is}
\begin{equation}
u(x,y) = 1-\frac{|d_e(\phi(x), \phi(t))-d_e(\phi(y), \phi(t))|}{d_e(\phi(t), \hat{\phi}(t))}.\label{eq19}
\end{equation}

With~\autoref{eq15} and~\autoref{eq17}, it holds that:
\begin{equation}\label{eq20}
    0~\leq\frac{|d_e(\phi(x), \phi(t))-d_e(\phi(y), \phi(t))|}{d_e(\phi(t), \hat{\phi}(t))}<~1~.
\end{equation}

With~\autoref{eq19} and~\autoref{eq20}, it can be deduced that:
\begin{equation}\label{eqsu}
0~<u(x,y)\leq~1~\text{and}~\Delta u=1.\\
\end{equation}

Given a privacy parameter $\epsilon\geq0$, the probability of obtaining an output of the perturbed token $y \in C\!_r(t)$ for any input token $x \in C\!_r(t)$ is as follows:

\begin{align}
&Pr[y|x] = \frac{\exp\left(\frac{\epsilon \cdot u(x,y)}{2\Delta u}\right)}{\sum_{y'\in C\!_r(t)} \exp\left(\frac{\epsilon \cdot u(x,y')}{2\Delta u}\right)}\label{p_sample}\\
&= \frac{\exp\left(\frac{\epsilon}{2}\cdot\left(1 - \frac{|d_e(\phi(x), \phi(t)) - d_e(\phi(y), \phi(t))|}{d_e(\phi(t), \hat{\phi}(t))}\right)\right)}{\sum_{y' \in C\!_r(t)} \exp\left(\frac{\epsilon}{2}\cdot\left(1 - \frac{|d_e(\phi(x), \phi(t)) - d_e(\phi(y'), \phi(t))|}{d_e(\phi(t), \hat{\phi}(t))}\right)\right)}.
\end{align}

Specifically for the input token $t \in C\!_r(t)$ and output token $y \in C\!_r(t)$,  it can be deduced that:
\begin{align}
    u(t,y) &=1-\frac{d_e(\phi(y), \phi(t))}{d_e(\phi(t), \hat{\phi}(t))},
\end{align}
\begin{align}
&Pr[y|t]= \frac{\exp\left(\frac{\epsilon}{2}\cdot\left(1 - \frac{d_e(\phi(t), \phi(y))}{d_e(\phi(t), \hat{\phi}(t))}\right)\right)}{\sum_{y' \in C\!_r(t)} \exp\left(\frac{\epsilon}{2}\cdot\left(1 - \frac{d_e(\phi(t), \phi(y'))}{d_e(\phi(t), \hat{\phi}(t))}\right)\right)}.
\end{align}

The detailed process of RANTEXT is shown in~\autoref{alg:RANTEXT}. Furthermore, the token sampling for each raw token in RANTEXT satisfies the definition of \(\epsilon\)-LDP:

\begin{theorem}\label{theorem:3}
Given a privacy parameter \(\epsilon \geq 0\) and a \textit{random adjacency list} \(C\!_r(t)\) of token \(t\), for any input tokens \(x, x' \in C\!_r(t)\) and output token \(y \in C\!_r(t)\), the randomized mechanism \(M\) of RANTEXT holds that:
\begin{equation}
\frac{\Pr[M(x) = y]}{\Pr[M(x') = y]} \leq e^\epsilon .
\end{equation}
\end{theorem}
\autoref{theorem:3} demonstrates that given a \(C\!_r(t)\) of token $t$, the token sampling for each raw token in RANTEXT satisfies \(\epsilon\)-LDP.~\autoref{theorem:3} is proven in~\hyperref[app:proof]{Appendix C}.

\section{Experiments}\label{sec:exp}
In this section, we evaluate the privacy protection levels and utility of the \Name with various LDP mechanisms.

\begin{figure*}[t!]
%\centering
    % 调整图片间的空白间隔，根据实际需要进行调整
    \hspace{10pt}
    \setlength{\tabcolsep}{-8pt} 
    \centering
    \begin{tabular}{ccc}
    \subfloat[\small CNN/Daily Mail ]{
        \includegraphics[width=0.33\textwidth]{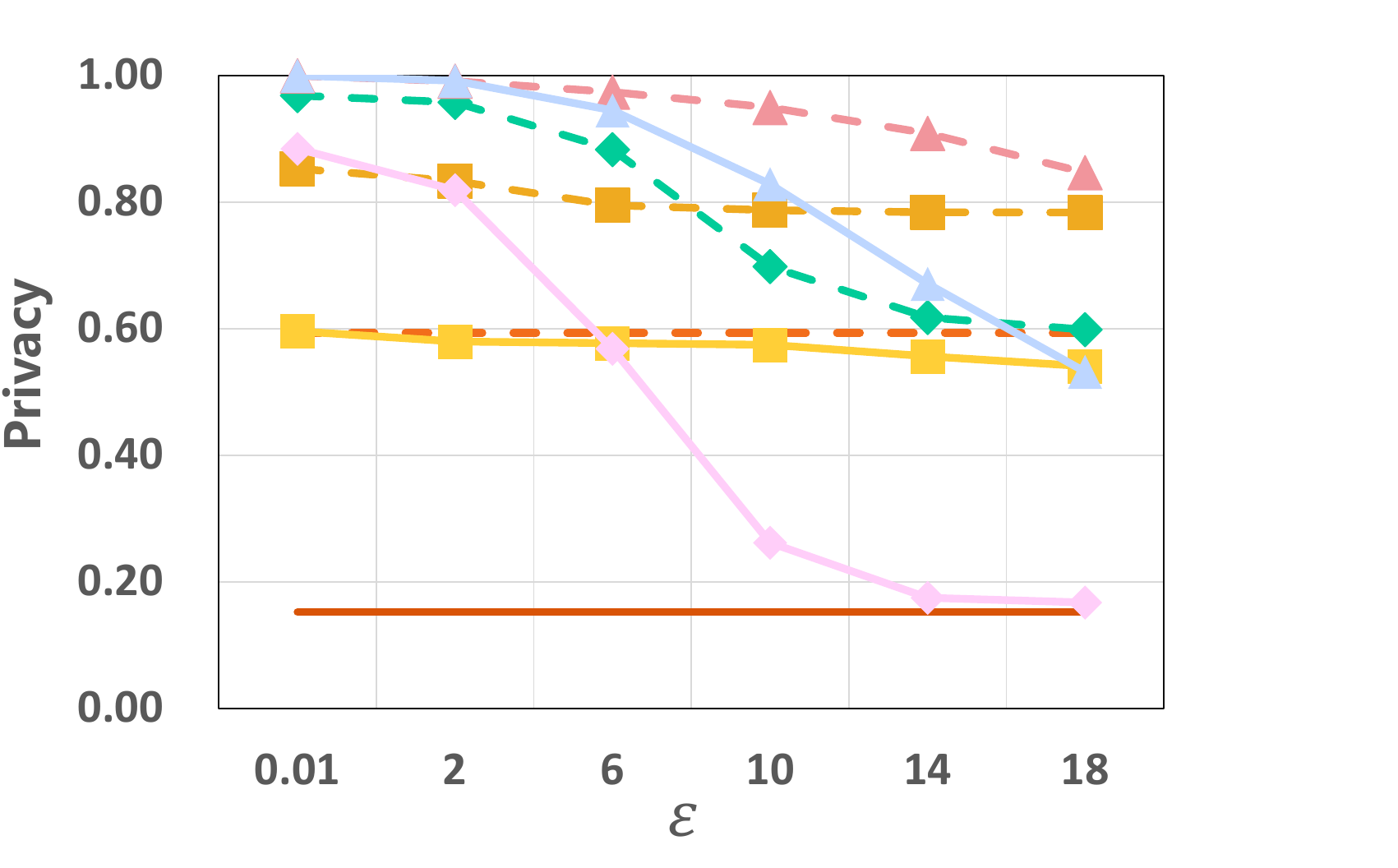}
    } &
    \subfloat[\small Wikitext-103-v1  ]{
        \includegraphics[width=0.33\textwidth]{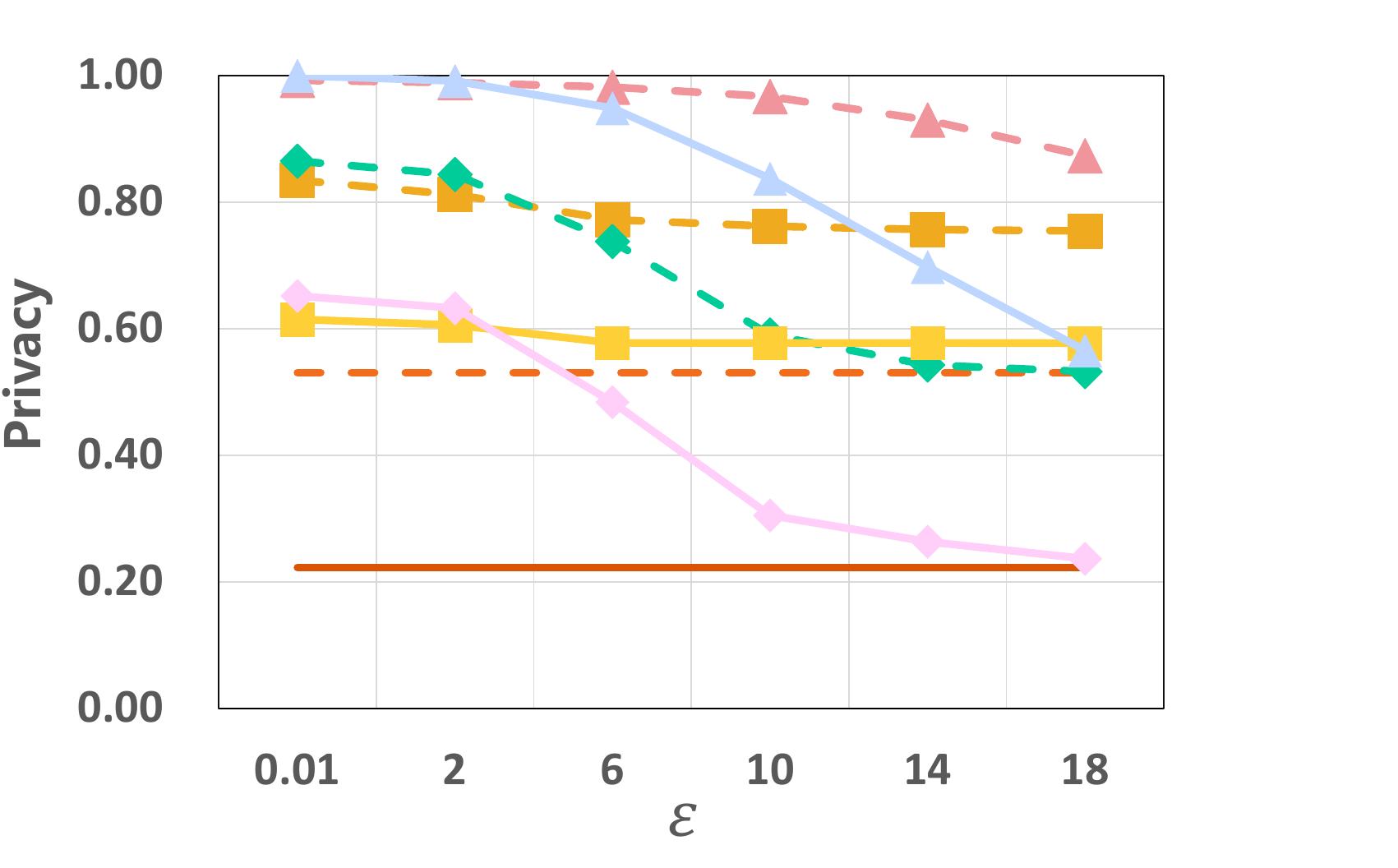}
    } &
    \subfloat[\small ArXiv Dataset ]{
        \includegraphics[width=0.33\textwidth]{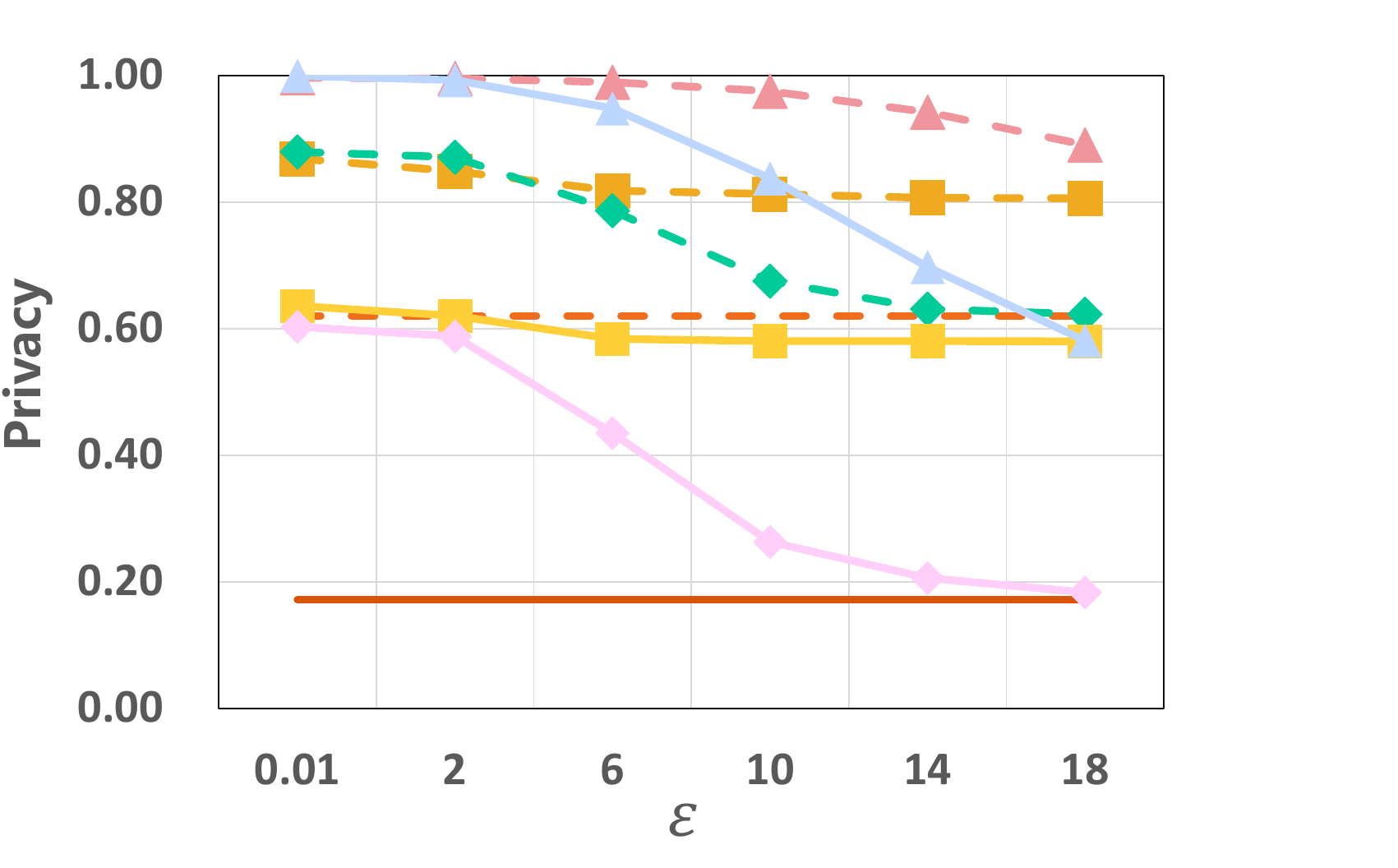}
    }
    \end{tabular}\\
    \centering
    \vspace{2pt}
    \includegraphics[width=0.53\textwidth]{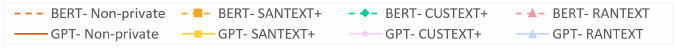}
    \caption{Results of BERT inference attack and GPT inference attack on CNN/Daily Mail, Wikitext-103-v1, and ArXiv Dataset.}
    \label{fig:BERT}
    \vspace{-10pt}
\end{figure*}
\begin{figure*}[t!]
    % 调整图片间的空白间隔，根据实际需要进行调整
    \hspace{10pt}
    \setlength{\tabcolsep}{-8pt} 
    \centering
    \begin{tabular}{ccc}

    \subfloat[\small CNN/Daily Mail]{
        \includegraphics[width=0.33\textwidth]{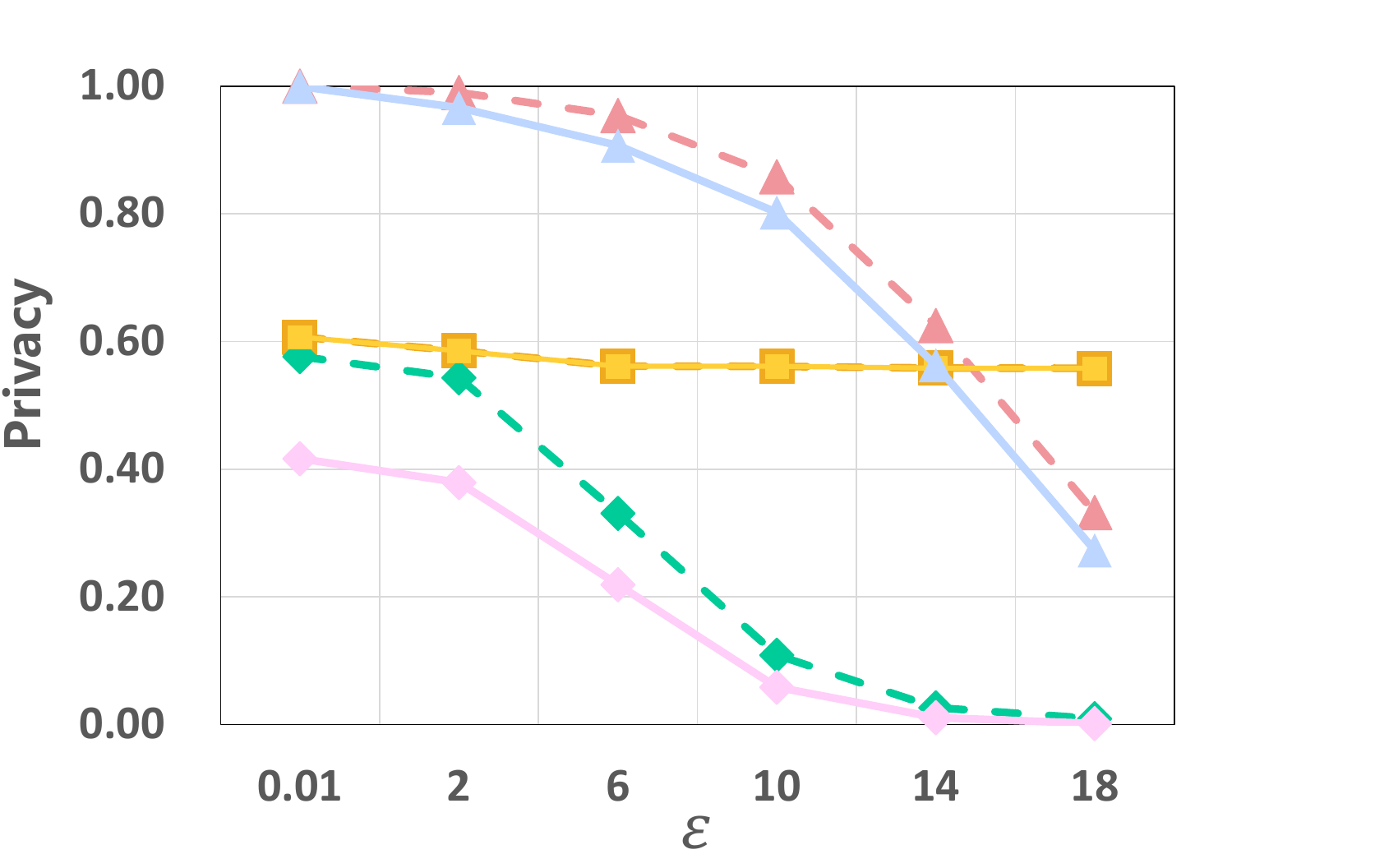}
    } &
    \subfloat[\small Wikitext-103-v1]{
        \includegraphics[width=0.33\textwidth]{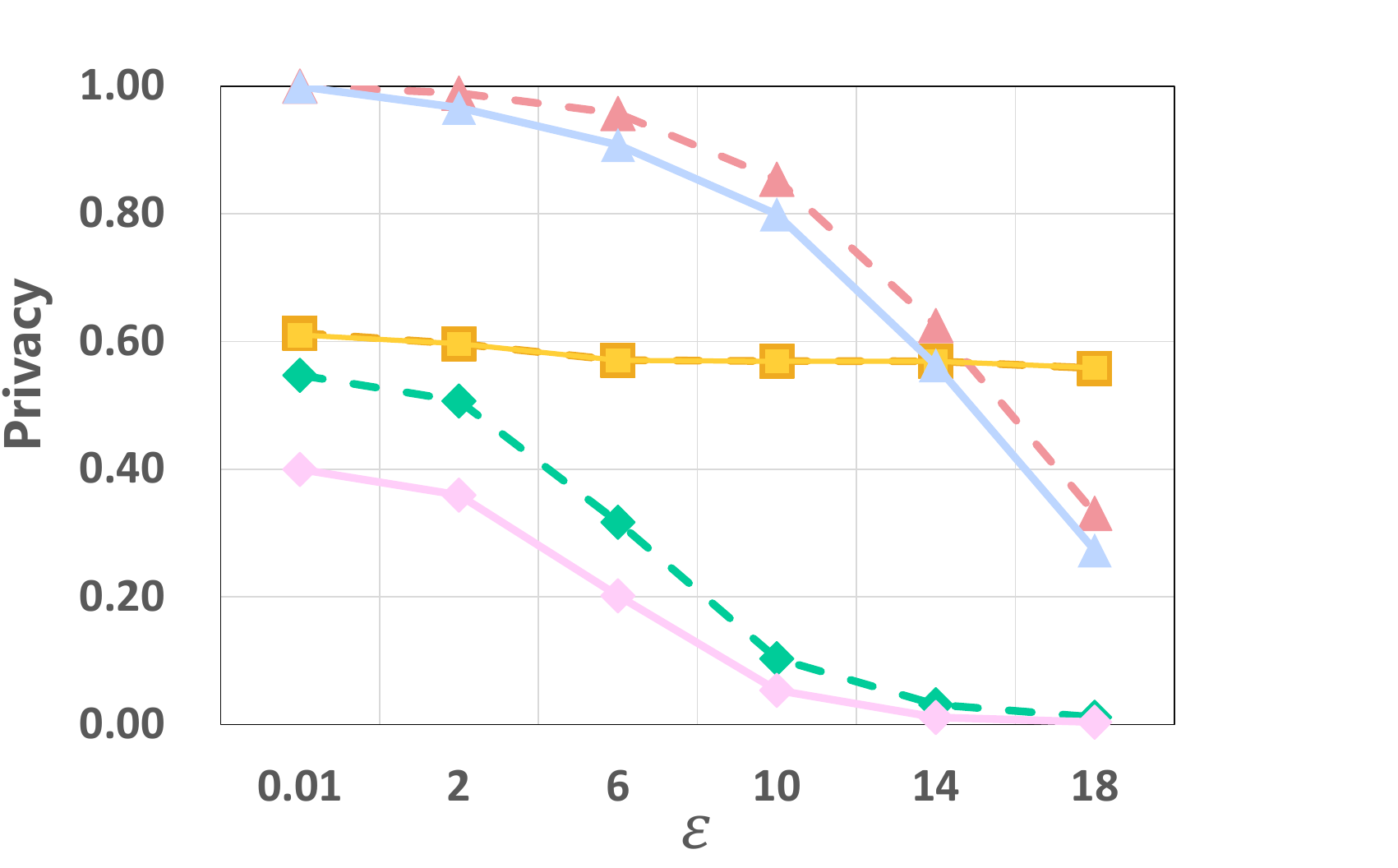}
    } &
    \subfloat[\small ArXiv Dataset ]{
        \includegraphics[width=0.33\textwidth]{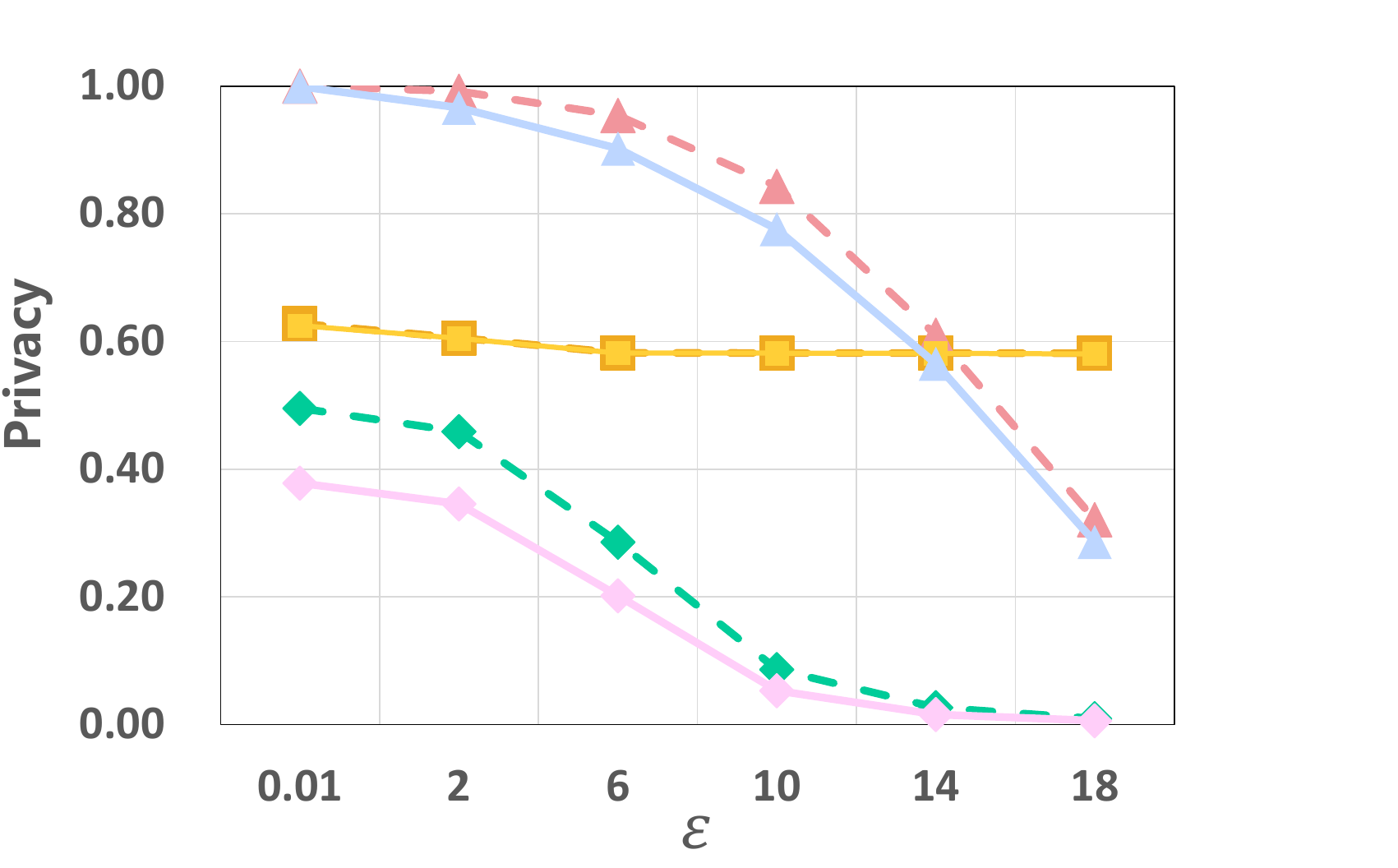}
    }
    \end{tabular}
    \\
    \centering
    \includegraphics[width=0.74\textwidth]{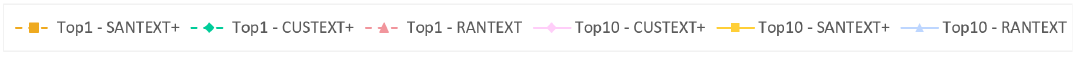}
    \caption{Results of embedding inversion attack on CNN/Daily Mail, Wikitext-103-v1 and ArXiv Dataset.}
    \label{fig:KNN250}
\end{figure*}

\subsection{Experiment Setup}
\mypara{Datasets} For traditional open-ended text generation tasks, we use two classic NLP datasets: CNN/Daily Mail~\cite{hermann2015teaching} and Wikitext-103-v1~\cite{merity2016pointer}. For practical applications, we use the ArXiv Dataset~\cite{clement2019arxiv}, FinRED~\cite{sharma2022finred}, and MedQA~\cite{jin2021disease}. These datasets encompass a wide range of events and entities.

\mypara{Baselines} \Name is the first practical framework for privacy-preserving inference that implements differential privacy in text generation tasks~\cite{ xu2022learning}. As there are no other frameworks of the same type, we did not compare \Name with any others. For the differentially private mechanisms of the perturbation module, we compared RANTEXT with existing state-of-the-art mechanisms, SANTEXT+~\cite{yue-etal-2021-differential} and CUSTEXT+~\cite{chen2023customized} in the default settings of them.

\mypara{Implementation} 
We conduct experiments on a cluster equipped with NVIDIA RTX A6000 GPUs and Intel Xeon Gold 6130 2.10 GHz CPUs. We use GPT-4~\cite{gpt4} as the remote large language model. Its token vocabulary is cl100k\_base~\cite{cl100k_base}, from which we select the first 11,000 English tokens as $V$. The embedding function $\phi(\cdot)$ is text-embedding-ada-002~\cite{text-embedding-ada-002}. For the local extraction module, we employ Vicuna-7b-4bit~\cite{zheng2023judging}, Llama2-7b-4bit~\cite{touvron2023llama}, and Llama3.1-8b-4bit~\cite{touvron2023llama}.

\begin{figure*}[t]\label{fig:sim_l}
    \setlength{\tabcolsep}{-8pt} 
    \hspace{10pt}
    \begin{tabular}{ccc}
    \subfloat[\small  CNN/Daily Mail ]{
        \includegraphics[width=0.32\textwidth]{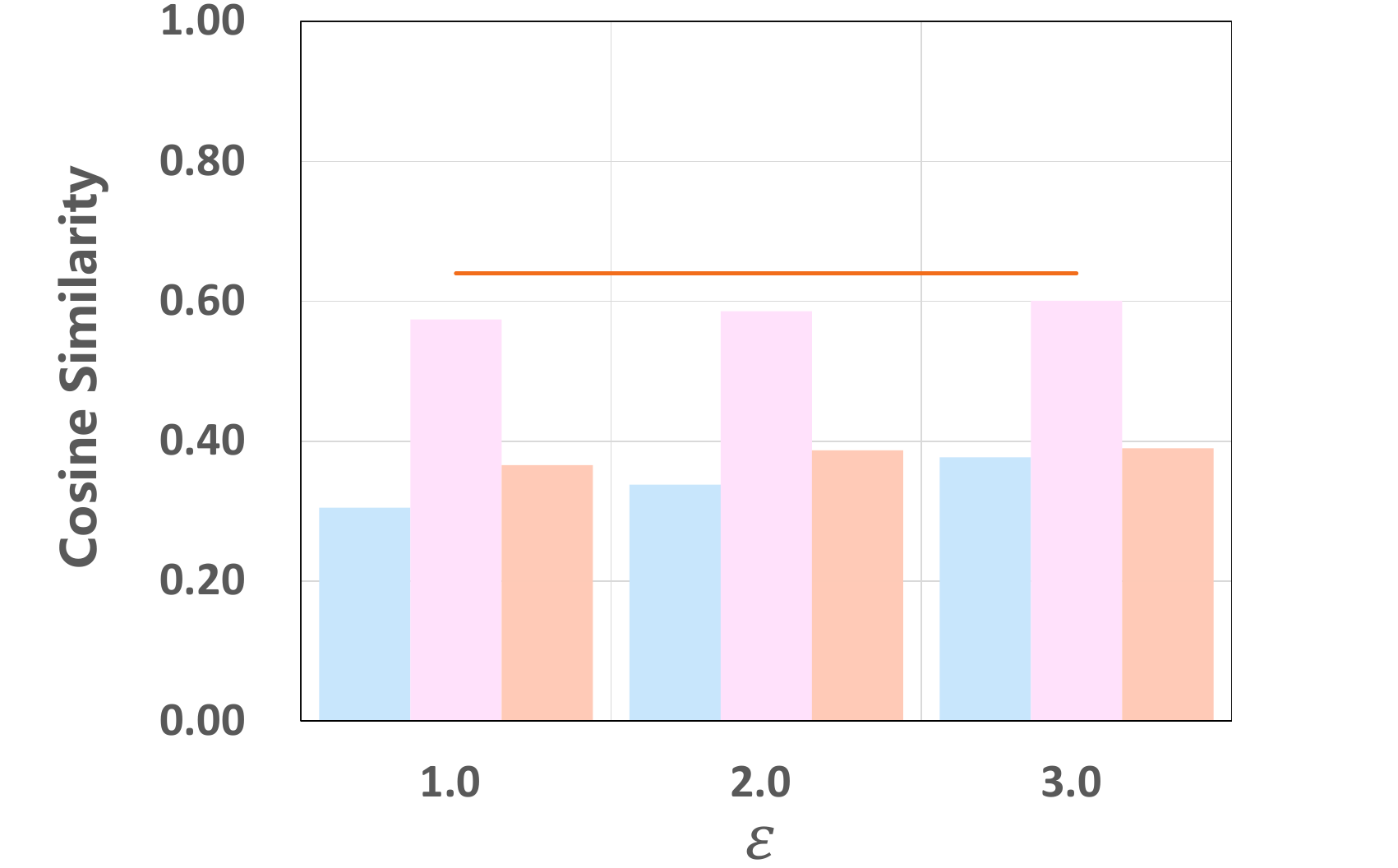}
    } &
    \subfloat[\small  Wikitext-103-v1]{
        \includegraphics[width=0.32\textwidth]{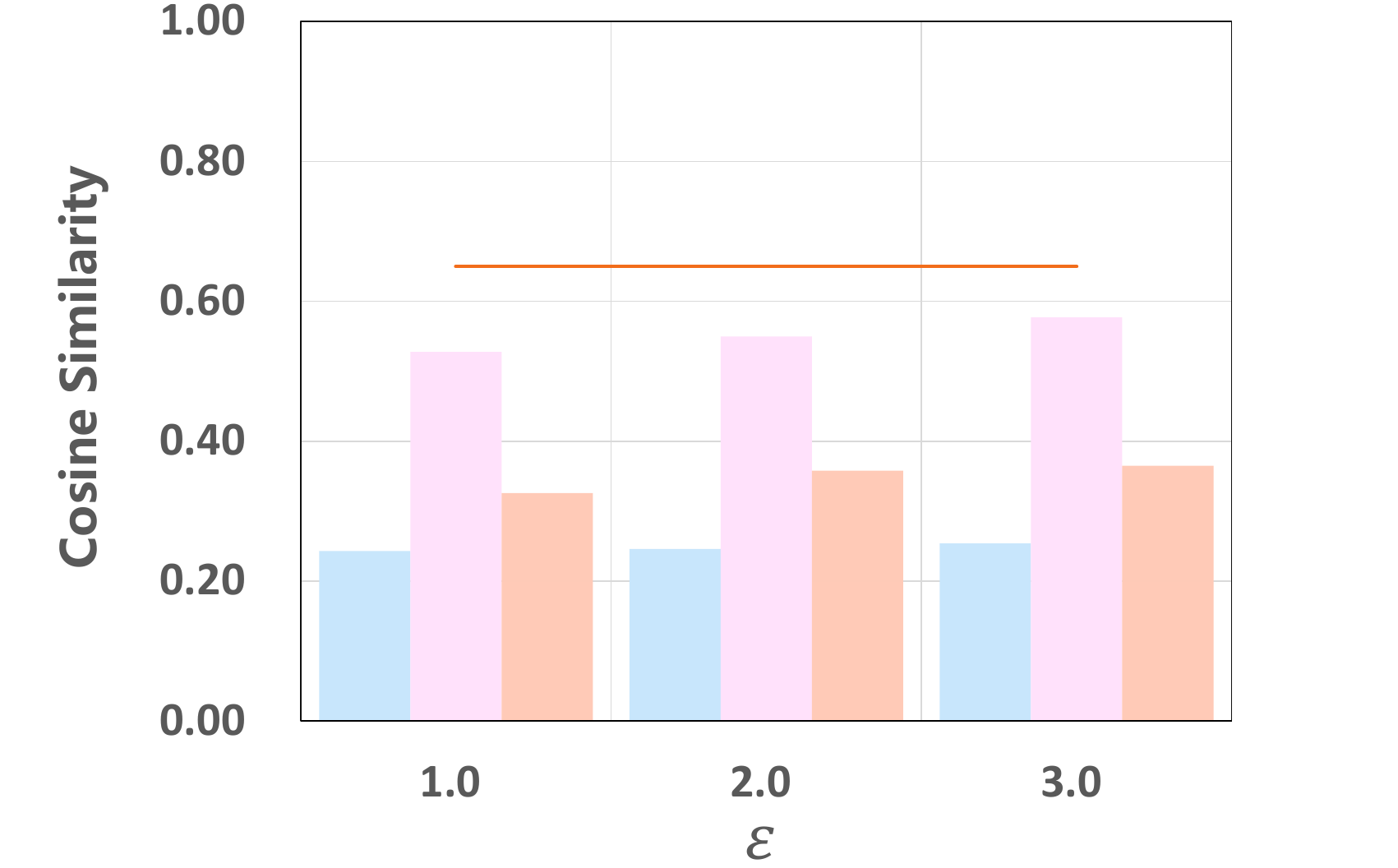}
    } &
    \subfloat[\small  ArXiv Dataset]{
        \includegraphics[width=0.32\textwidth]{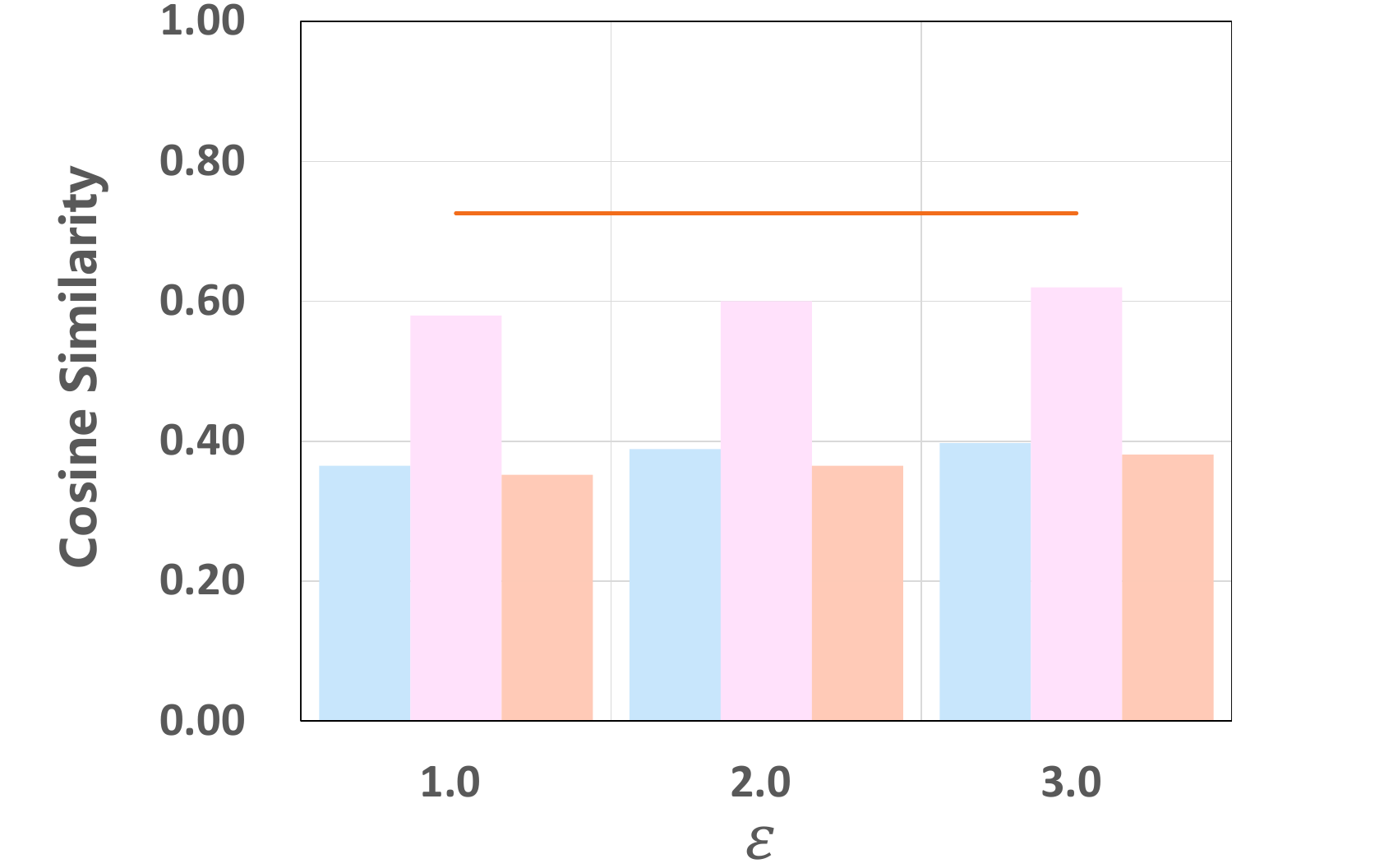}
    }
    \end{tabular}\\
    \centering
    \includegraphics[width=0.46\textwidth]{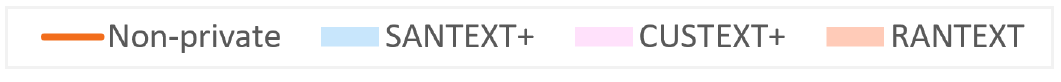}
    \caption{Cosine similarity $\downarrow$ between the perturbed generation and the raw document.}
    \label{fig:leak_gen}
\end{figure*}
\begin{table}[t]
\centering
\caption{Privacy leakage via the perturbed generation.}
\label{tbl:pril}
\vspace{-2pt}
 \resizebox{0.78\columnwidth}{!}{
\begin{tabular}{c|l||ccc}
\toprule
\multirow{2}{*}{\textbf{$\epsilon$}} & \multirow{2}{*}{\textbf{Method}} 
& \multicolumn{3}{c}{\textbf{Privacy Leakage Rate\,$\downarrow$}} \\ \cline{3-5} 
 & 
\rule{0pt}{2.5ex} & 1-gram & 2-gram  & 3-gram\\
 \midrule
 \multirow{1}{*}{$\infty$}   & Non-private
 & $0.110$ & $0.058$ & $0.027$ \\
 \midrule
 \multirow{3}{*}{$2.0$} & SANTEXT$^+$
& $0.090$ & $0.030$ & $0.016$ \\ 
 & CUSTEXT$^+$
& $0.098$ & $0.030$ & $0.007$ \\ 
 & RANTEXT
& $\mathbf{0.078}$ & $\mathbf{0.024}$ & $\mathbf{0.004}$ \\ 
 \midrule
 \multirow{3}{*}{$6.0$} & SANTEXT$^+$
& $0.103$ & $0.053$ & $0.022$ \\ 
 & CUSTEXT$^+$ 
& $0.099$ & $0.034$ & $0.008$ \\ 
 & RANTEXT  
& $\mathbf{0.081}$ & $\mathbf{0.027}$ & $\mathbf{0.005}$ \\ 
 \midrule
 \multirow{3}{*}{$10.0$} & SANTEXT$^+$
& $0.104$ & $0.055$ & $0.022$ \\ 
 & CUSTEXT$^+$  
& $0.103$ & $0.04$ & $0.013$ \\ 
 & RANTEXT 
& $\mathbf{0.095}$ & $\mathbf{0.029}$ & $\mathbf{0.006}$ \\ 
 \midrule
 \multirow{3}{*}{$14.0$} & SANTEXT$^+$
& $0.105$ & $0.057$ & $0.023$ \\ 
 & CUSTEXT$^+$  
& $0.105$ & $0.052$ & $0.017$ \\ 
 & RANTEXT 
& $\mathbf{0.101}$ & $\mathbf{0.032}$ & $\mathbf{0.008}$ \\ 
 \bottomrule
\end{tabular}
}
\vspace{-6pt}
\end{table}

\subsection{BERT Inference Attack}\label{sec:privacy}
In the \textit{BERT inference attack}~\cite{yue-etal-2021-differential}, an adversary employs a pre-trained BERT model to recover raw document $Doc$ from their perturbed version $Doc_p$. 
The BERT model, developed through {masked language modeling}~\cite{bao2020unilmv2}, predicts the raw tokens by sequentially replacing each token in the perturbed text with a special token \texttt{"[MASK]"}. This approach leverages BERT's capability to understand context, allowing it to infer the masked tokens. An attack is successful if the output token matches the input token. Subsequently, we calculate attack success rate\footnote{We exclude \texttt{"[UNK]"} token, as it does not yield meaningful information.} across all attacks, denoted as $ r\!_{{ats}} $. The privacy protection level of the differentially private mechanism is defined as $ 1 - r\!_{{ats}}$.

As shown in~\autoref{fig:BERT}, RANTEXT offers better privacy protection against \textit{BERT inference attack} compared to SANTEXT+ and CUSTEXT+. The experimental results indicate that RANTEXT provides over 80\% privacy protection within an \( \epsilon \) value range of 0.01 to 18.0. In particular, with an $\epsilon$ value of 18.0 on the CNN/Daily Mail dataset, the privacy protection level of RANTEXT is 1.11 $\times$ that of SANTEXT+ and 1.41 $\times$ that of CUSTEXT+. We analyzed the results of the experiment and found that BERT did not recognize the tokens of GPT-4. To more comprehensively evaluate RANTEXT's security, we proposed an adaptive attack leveraging the capabilities of GPT-4 in~\autoref{sec:p_gen}, GPT inference attack. 
\subsection{Embedding Inversion Attack}\label{emdatt}
\textit{Embedding inversion attack}~\cite{qu2021natural} computes the distance between the embedding of each token in the perturbed document and the embeddings of other tokens in the vocabulary, returning top $K$ tokens with the closest Euclidean distance. The privacy protection level is defined as $ 1 - r\!_{{ats}}$.

Experiments were conducted under the conditions of top $K=1$ and $10$.~\autoref{fig:KNN250} illustrates that, under both conditions, SANTEXT+ and CUSTEXT+ are susceptible to embedding inversion attacks, indicating a relatively lower level of privacy protection. Even at $\epsilon=0.01$, these methods could only provide privacy protection for over 40\% of the original documents. As the top $K$ changes from $1$ to $10$, the privacy protection level of SANTEXT+ and CUSTEXT+ remains largely unchanged. On the other hand, RANTEXT benefits from its design of the \textit{random adjacency list} (generally larger than that in CUSTEXT+) and the perturbation on each token, preventing attackers from successfully reconstructing raw tokens.

\subsection{Adaptive Attack: GPT Inference Attack}
\label{sec:p_gen}

RANTEXT applies perturbations to the GPT-4 token vocabulary. Since GPT-4 recognizes all tokens, it is hypothesized that GPT-4 can better reconstruct raw tokens perturbed by RANTEXT. Therefore, we propose an adaptive attack, the \textit{GPT inference attack}. In this method, the attacker inputs perturbed text into GPT-4 and instructs it to recover each token. The attack is successful if the recovered token coincides with the raw one. The privacy protection level is defined as $ 1 - r\!_{{ats}}$.
The prompt of this attack can be found in \hyperref[app:C]{Appendix D}.

\begin{table*}[t]
\centering
\caption{Performance comparison on open-ended text generation tasks across different methods, datasets, and privacy parameters ($\epsilon = 1, 2, 3$), evaluated based on diversity, MAUVE, and coherence.}
\label{tab:utility-result}
\resizebox{0.93\textwidth}{!}{
\begin{tabular}{l|l||c c c|c c c|c c c}
\toprule
\multirow{2}{*}{\textbf{Dataset}} & \multirow{2}{*}{\textbf{Method}}  & \multicolumn{3}{c|}{\textbf{diversity$\uparrow $ }} & \multicolumn{3}{c|}{\textbf{MAUVE$\uparrow $ }} & \multicolumn{3}{c}{\textbf{coherence$\uparrow $ }} \\ 
\cline{3-11}
   \rule{0pt}{2.3ex}                      &                          & $\epsilon=1.0$ & $\epsilon=2.0$ & $\epsilon=3.0$ & $\epsilon=1.0$ & $\epsilon=2.0$ & $\epsilon=3.0$ & $\epsilon=1.0$ & $\epsilon=2.0$ & $\epsilon=3.0$ \\ 
\midrule
\multirow{6}{*}{\textbf{CNN/Daily Mail}} & GPT-4 & \multicolumn{3}{c|}{$0.983$} & \multicolumn{3}{c|}{$0.671$} & \multicolumn{3}{c}{$0.632$} \\
& Vicuna-7b-4bit (3.89GB) & \multicolumn{3}{c|}{$0.943$} & \multicolumn{3}{c|}{$0.197$} & \multicolumn{3}{c}{$0.627$} \\
\cline{2-11}
 \rule{0pt}{2.5ex}
& \Name+ SANTEXT$^+$ & $0.966$ & $0.967$ & $0.966$ & $0.351$ & $0.374$ & $0.407$ & $0.590$ & $0.632$ & $0.642$ \\
& \Name+ CUSTEXT$^+$ & $0.966$ & $0.967$ & $0.965$ & $0.540$ & $\mathbf{0.571}$ & ${0.581}$ & $\mathbf{0.726}$ & $0.733$ & $\mathbf{0.752}$ \\
& \Name+ RANTEXT & $\mathbf{0.970}$ & $\mathbf{0.970}$ & $\mathbf{0.971}$ & $\mathbf{0.542}$ & $0.563$ & $\mathbf{0.587}$ & $0.723$ & $\mathbf{0.735}$ & $0.736$ \\
\midrule
\multirow{6}{*}{\textbf{Wikitext-103-v1}} & GPT-4 & \multicolumn{3}{c|}{$0.987$} & \multicolumn{3}{c|}{$0.453$} & \multicolumn{3}{c}{$0.672$} \\
& Vicuna-7b-4bit (3.89GB) & \multicolumn{3}{c|}{$0.916$} & \multicolumn{3}{c|}{$0.158$} & \multicolumn{3}{c}{$0.663$} \\
\cline{2-11}
 \rule{0pt}{2.5ex}
& \Name+ SANTEXT$^+$ & $0.958$ & $0.958$ & $0.959$ & $0.213$ & $0.220$ & $0.255$ & $0.650$ & $0.658$ & $0.678$ \\
& \Name+ CUSTEXT$^+$ & $0.960$ & $0.961$ & $0.959$ & $\mathbf{0.301}$ & $\mathbf{0.315}$ & $\mathbf{0.321}$ &$ 0.727$ & $0.736$ & ${0.741}$ \\
& \Name+ RANTEXT & $\mathbf{0.961}$ & $\mathbf{0.962}$ & $\mathbf{0.961}$ & $0.245$ & $0.254$ & $0.274$ & $\mathbf{0.729}$ & $\mathbf{0.744}$ & $\mathbf{0.745}$ \\
\midrule
\multirow{6}{*}{\textbf{ArXiv Dataset}} & GPT-4 & \multicolumn{3}{c|}{$0.935$} & \multicolumn{3}{c|}{$0.736$} & \multicolumn{3}{c}{$0.726$} \\

& Vicuna-7b-4bit (3.89GB) & \multicolumn{3}{c|}{$0.873$} & \multicolumn{3}{c|}{$0.366$} & \multicolumn{3}{c}{$0.703$} \\
\cline{2-11}
 \rule{0pt}{2.5ex}
& \Name+ SANTEXT$^+$ & $0.945$ & ${0.946}$ & $0.946$ & $0.196$ & $0.207$ & $0.230$ & $0.651$ & $0.670$ & $0.690$ \\
& \Name+ CUSTEXT$^+$ & $0.946$ & $0.945$ & $0.944$ & $\mathbf{0.410}$ & $\mathbf{0.443}$ & $\mathbf{0.455}$ & $0.748$ & $\mathbf{0.767}$ & $\mathbf{0.784}$ \\
& \Name+ RANTEXT & $\mathbf{0.947}$ & $\mathbf{0.948}$ & $\mathbf{0.947}$ & $0.359$ & $0.375$ & $0.395$ & $\mathbf{0.752}$ & $0.761$ & $0.762$ \\
\bottomrule
\end{tabular}
}
\end{table*}

\begin{table}[t]
\centering
\caption{Performance Comparison of the Time Cost Per Inference on 100 tokens in \Name.}
\label{tbl:time}
\vspace{-2pt}
 \resizebox{0.999\columnwidth}{!}{
\begin{tabular}{l||ccc}
\toprule
\multirow{2}{*}{\textbf{Method}} 
& \multicolumn{3}{c}{\textbf{Time Cost (seconds)
}} \\ \cline{2-4} 
 & 
\rule{0pt}{2.5ex}  SANTEXT$^+$ &CUSTEXT$^+$ & RANTEXT\\
 \midrule
Perturbation Module
& $0.0015\pm0.0001$ &${0.0005\pm0.0001}$  &${0.0543\pm0.0023}$  \\ 

  Black-box Inference
&-  &$2.8324\pm0.2111$ &-  \\ 
 Extraction Module
&-  &$3.5673\pm0.2781$ &-  \\ 
 \bottomrule
\end{tabular}
}
\vspace{-6pt}
\end{table}

\autoref{fig:BERT} displays the results of the GPT inference attack. GPT-4 has a higher attack success rate than BERT in all tests. This may be due to GPT-4's larger size and better understanding abilities, making it more effective in inference attacks. Confronted with the GPT inference attack, SANTEXT+ and CUSTEXT+ showed lower privacy protection levels than RANTEXT, which maintained the best privacy protection.

\subsection{Privacy Leakage in Perturbed Generation}

We further discussed the possibility of the raw document $Doc$ being leaked by the perturbed generation result $Gen_p$. \autoref{fig:leak_gen} shows the cosine similarity between $Doc$ and $Gen_p$. The orange straight line indicates the cosine similarity between $Doc$ and the generation of GPT-4 without any privacy protection. Experimental results reveal that RANTEXT maintains low semantic similarity between the raw document $Doc$ and the perturbed generation result $Gen_p$, indicating the low risk of privacy leakage through perturbed results.

Moreover, we measured privacy leakage in perturbed outputs by checking if n-gram tokens from the original document were repeated. A n-gram token found in both raw text and perturbed output counts as a leak. As \autoref{tbl:pril} shows, even with non-private prompts, under 11\% privacy of raw document is leaked.

\subsection{Evaluation of Utility}\label{sec:utility}
%We assessed output quality generated by \Name using various differentially private mechanisms within the perturbation module on different datasets when the local model was set to Vicuna-7b-4bit (3.89GB) within the extraction module.~\autoref{tab:utility-result} presents the main results of \Name's generation quality compared to those of non-private GPT-4. From the experimental results, we observe:

We evaluated the quality of outputs generated by \Name with various differentially private mechanisms in the perturbation module, using the Vicuna-7b-4bit (3.89GB) in the extraction module on various datasets. Following previous works of open-ended text generation~\cite{welleck2019neural, xu2022learning}, we use the first 50 tokens of the articles referred to raw document $Doc$, which we must protect. We use the continuation writing of $Doc$  referred to as $Gen$, which consists of 100 tokens. Tokens are counted by the tokenizer function of GPT-2~\cite{lagler2013gpt2}. Aligning with~\cite{lin2021straight}, three metrics were employed to evaluate the quality of the generated text in the open-ended generation task:\\
\!\!\!1) \textit{Diversity.} This metric suggests the text's diversity by computing the unique n-gram repetition rates as follows:\[ diversity = \sum_{n=2}^{4} \frac{\left|{unique~n\!-\!grams}(Gen)\right|}{\left|{total~n\!-\!grams}(Gen)\right|}.\]
A lower score indicates that the model is prone to repetition, while a higher score shows broader vocabulary usage.\\
2) \textit{MAUVE~\textup{\cite{pillutla2021mauve}}.} It is employed to assess the similarity between text generated by a language model and human-authored target continuation text. A higher score is desirable in this metric.\\
3) \textit{Coherence.} Coherence computes the cosine similarity between embeddings of document $Doc$ and continuation $Gen$: 
\begin{equation*}
COH(Doc, Gen)\!=\!\frac{\text{SimCSE}(Doc)\!\cdot\!\text{SimCSE}(Gen)}{\|\text{SimCSE}(Doc)\|\!\cdot\! \|\text{SimCSE}(Gen)\|},
\end{equation*}
\text{where}  \text{SimCSE}(x) \text{ represents the pretrained model~\cite{gao2021simcse}}.

\autoref{tab:utility-result} shows \Name's generation quality compared to non-private GPT-4:

(1) Although the uploaded prompt is perturbed by differential privacy, the quality of text generated by \Name is comparable to that directly produced by non-private GPT-4 and better than the local model's output. It proves that \Name works effectively. (2) In terms of diversity, the quality of text generated by RANTEXT is superior to that of CUSTEXT+ and SANTEXT+. This phenomenon can be attributed to the design of the \textit{random adjacency list} $C\!_r$ in RANTEXT, which perturbs tokens to the more probable new ones without retaining them. However, in some specific topics, the variety of tokens is not particularly rich. Additionally, RANTEXT discards proper nouns (those not belonging to $V$) for privacy protection. As a result, RANTEXT's performance is slightly inferior to that of CUSTEXT+ with respect to MAUVE.
(3) From the perspective of coherence, experimental results indicate that RANTEXT and CUSTEXT+ outperform SANTEXT+. This is likely because SANTEXT+ uses the entire vocabulary as its \textit{static adjacency list}, which is too large for the utility of the perturbed text.

%With the same privacy parameter $\epsilon$, the quality of generated text using RANTEXT in \Name is equivalent to that of CUSTEXT+ and superior to that of SANTEXT+.
For practical deployment, we measured the time cost per inference in \Name. As illustrated in \autoref{tbl:time}, experimental results indicate that \Name does not require a significant amount of time. Most of the additional time is spent in the extraction module, which is less than 4 seconds.

We also investigated whether \Name works in different local models. \autoref{tbl:diff_size} demonstrate that \Name works well with different models and various privacy parameter $\epsilon$. We further compared the cosine similarity between the final generation of \Name and the output generated by GPT-4 without any privacy protection. The result of the comparison is depicted in~\autoref{tbl:seq_MIA}. 
Under the same privacy parameter $\epsilon$ across three datasets, the perturbed generation of RANTEXT generally exhibits cosine similarity values close to that of the best-performing CUSTEXT+. This is likely because RANTEXT discards proper nouns (those not belonging to $V$) in $Doc$, whereas CUSTEXT+ retains all of this key information without perturbation. This phenomenon is particularly evident in the Wikitext-103-v1 and ArXiv datasets, which contain more proper nouns. We emphasize that this is also one of the reasons why CUSTEXT+ is vulnerable to the embedding inversion attack.

\begin{figure*}[t]
\hspace{10pt}
\begin{minipage}{1\textwidth}
  \centering
  \includegraphics[width=\linewidth]{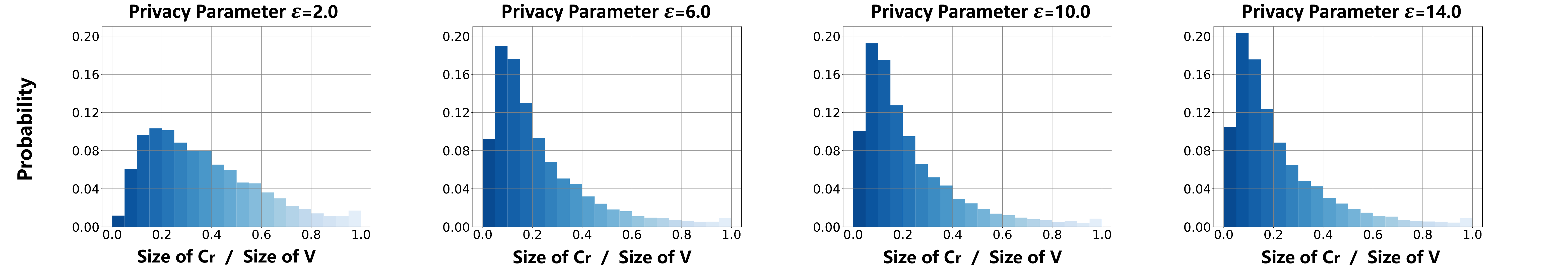}
  
\end{minipage}%
\caption{Probability distribution of the size of \textit{random adjacency list} under various $\epsilon$.}
    \label{fig:random_adjacent_proba}
\end{figure*}
\begin{table*}[t]
\centering
\caption{Comparison of the final generated text quality under different local models within the extraction module.}
\label{tbl:diff_size}
\resizebox{0.93\textwidth}{!}{
\begin{tabular}{l|l||c c c|c c c|c c c}
\toprule
\multirow{2}{*}{\textbf{Dataset}} & \multirow{2}{*}{\textbf{Method}}  & \multicolumn{3}{c|}{\textbf{diversity$\uparrow $ }} & \multicolumn{3}{c|}{\textbf{MAUVE$\uparrow $ }} & \multicolumn{3}{c}{\textbf{coherence$\uparrow $ }} \\ 
\cline{3-11}
   \rule{0pt}{2.3ex}                      &                          & $\epsilon=2.0$ & $\epsilon=6.0$ & $\epsilon=10.0$ & $\epsilon=2.0$ & $\epsilon=6.0$ & $\epsilon=10.0$ & $\epsilon=2.0$ & $\epsilon=6.0$ & $\epsilon=10.0$ \\ 
\midrule
\multirow{9}{*}{\textbf{CNN/Daily Mail}} & GPT-4 &&{$0.983$}& &&{$0.671$}& & &{$0.632$}& \\
& Llama2-7b-4bit (3.79GB) & &{$0.896$}& &&{$0.258$}& &&{$0.485$}& \\
& Vicuna-7b-4bit (3.89GB) & &{$0.943$}&  &&{$0.197$}&  &&{$0.627$}& \\
\cline{2-11}
 \rule{0pt}{2.5ex}
 & SANTEXT$^+$(Llama2-7b-4bit) & $0.964$ & $0.963$ & $0.962$ & $0.282$ & $0.374$ & $0.406$ & $0.226$ & $0.327$ & $0.342$ \\
& CUSTEXT$^+$(Llama2-7b-4bit) & $0.963$ & $0.962$ & $0.963$ & $\mathbf{0.493}$ & ${0.519}$ & ${0.548}$ & $\mathbf{0.460}$ & $\mathbf{0.483}$ & ${0.514}$ \\
& RANTEXT~(Llama2-7b-4bit) & $\mathbf{0.968}$ & $\mathbf{0.969}$ & $\mathbf{0.967}$ & ${0.473}$ & $\mathbf{0.526}$ & $\mathbf{0.566}$ & ${0.411}$ & ${0.453}$ & $\mathbf{0.525}$ \\
 \cline{2-11}
 \rule{0pt}{2.5ex}
& SANTEXT$^+$(Vicuna-7b-4bit) & $0.967$ & $0.969$ & $0.968$ & $0.374$ & $0.413$ & $0.448$ & $0.632$ & $0.679$ & $0.727$ \\
& CUSTEXT$^+$(Vicuna-7b-4bit) & $0.966$ & $0.967$ & $0.968$ & $\mathbf{0.571}$ & $\mathbf{0.632}$ & $\mathbf{0.670}$ & ${0.733}$ & ${0.749}$ & $\mathbf{0.789}$ \\
& RANTEXT~(Vicuna-7b-4bit) & $\mathbf{0.970}$ & $\mathbf{0.969}$ & $\mathbf{0.970}$ & ${0.563}$ & $0.586$ & ${0.635}$ & $\mathbf{0.735}$ & $\mathbf{0.753}$ & $0.773$ \\
\bottomrule
\end{tabular}
}
\end{table*}
\begin{table}[!t]
\centering
\caption{Cosine similarity$\uparrow$ between the final generation of \Name and the non-private generation from GPT-4.}
\label{tbl:seq_MIA}
\vspace{-2pt}
 \resizebox{0.90\columnwidth}{!}{
\begin{tabular}{l|l||ccc}
\toprule
\multirow{2}{*}{\textbf{Dataset}} &  \multirow{2}{*}{\textbf{Method}}
& \multicolumn{3}{c}{\textbf{$\epsilon$}} \\ \cline{3-5}
 & & \rule{0pt}{2.3ex}$1.0$ & $2.0$ & $3.0$\\
 \midrule
\multirow{3}{*}{\textbf{CNN/Daily Mail}} & SANTEXT$^+$
& $0.489$ & $0.499$ & $0.519$ \\
 & CUSTEXT$^+$
& ${0.571}$ & $\mathbf{0.579}$ &$\mathbf{0.585}$ \\ 
 & RANTEXT 
 & $\mathbf{0.574}$ & $\mathbf{0.579}$ &$0.584$ \\ 
 \midrule
\multirow{3}{*}{\textbf{Wikitext-103-v1}} & SANTEXT$^+$ 
& $0.544$ & $0.546$ &$0.572$ \\
 & CUSTEXT$^+$
& ${0.597}$ & $\mathbf{0.613}$ &$\mathbf{0.627}$ \\ 
 & RANTEXT 
 & $\mathbf{0.598}$ & ${0.609}$ &$0.617$ \\ 
 \midrule
\multirow{3}{*}{\textbf{ArXiv Dataset}} & SANTEXT$^+$ 
& $0.584$ & $0.591$ &$0.595$ \\
 & CUSTEXT$^+$
& $\mathbf{0.682}$ & $\mathbf{0.693}$ &$\mathbf{0.694}$ \\ 
 & RANTEXT 
 & ${0.655}$ & ${0.658}$ &$0.663$ \\ 
 \bottomrule
\end{tabular}
}
\vspace{-6pt}
\end{table}

\begin{figure*}[t]
%\centering
\hspace{4pt}
    % 调整图片间的空白间隔，根据实际需要进行调整
    \setlength{\tabcolsep}{-4pt} 
    \hspace{9.4pt}
    \begin{tabular}{ccc}

        \includegraphics[width=0.33\textwidth]{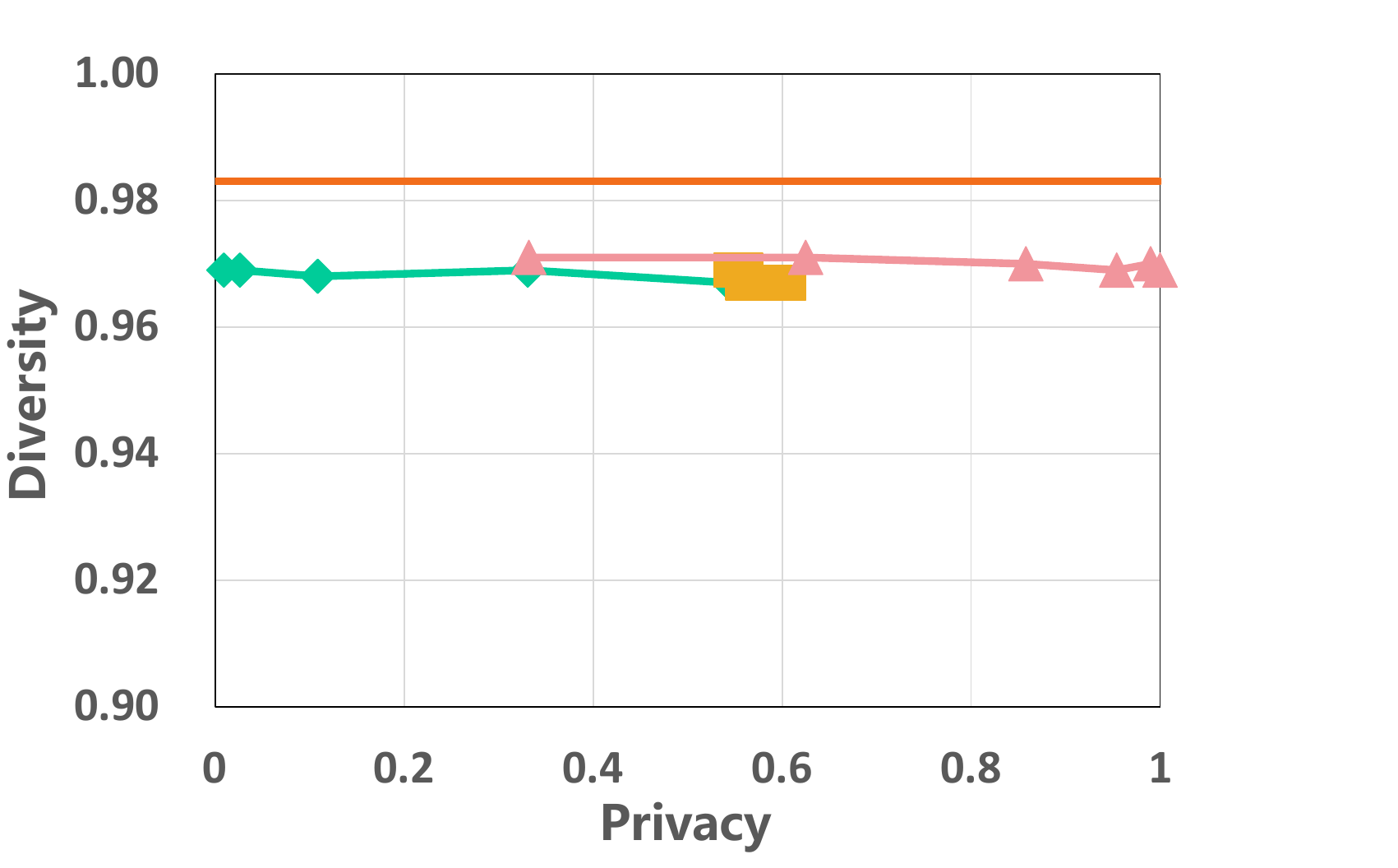}
     &
    
        \includegraphics[width=0.33\textwidth]{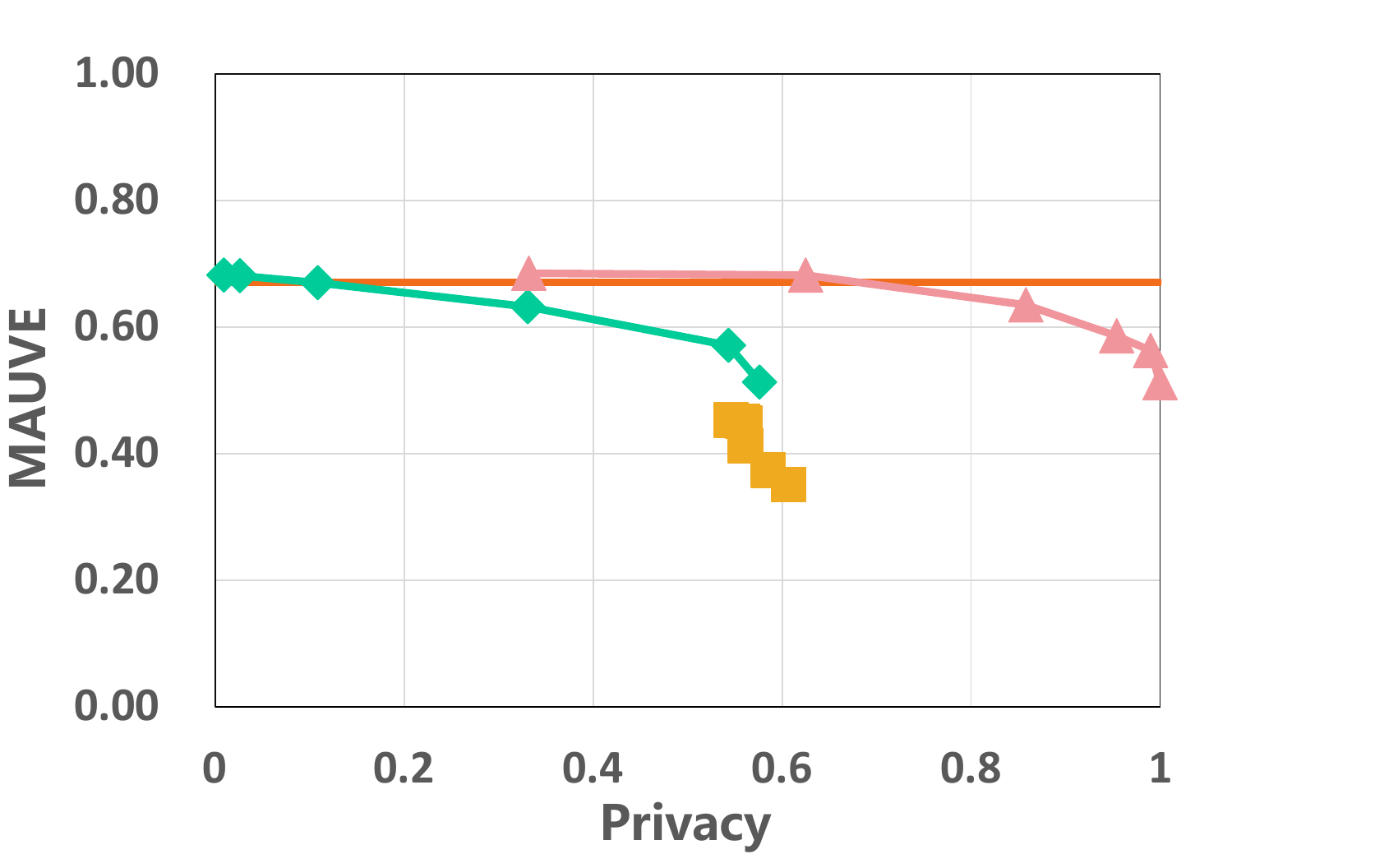}
    &
    \includegraphics[width=0.33\textwidth]{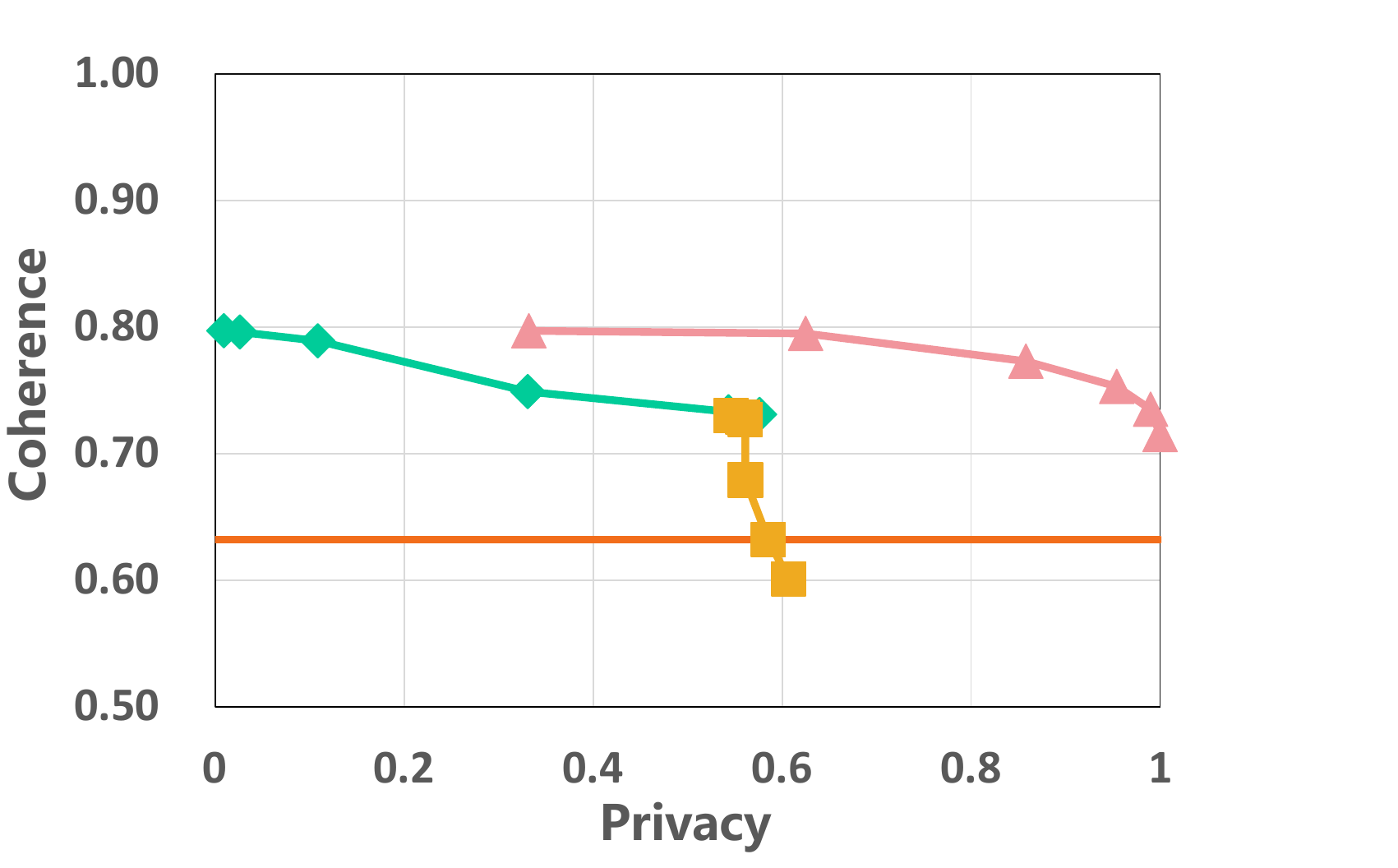}
    \end{tabular}\\
    \centering
    \includegraphics[width=0.47\textwidth]{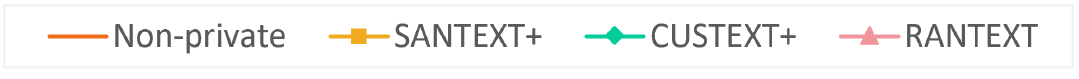}
    \caption{Results of the trade-off between utility and privacy protection with various privacy parameters $\epsilon$ ranging from 0.01 to 18.0.}
    \label{fig:trade-off}
\end{figure*}

Furthermore, we investigated the impact of the privacy parameter $\epsilon$ on the probability distribution of the size of the \textit{random adjacency list} $C\!_r$ in RANTEXT. We use $C\!_r/V$ to represent the proportion of $C\!_r$ in the entire vocabulary $V$. As shown in~\autoref{fig:random_adjacent_proba}, the \textit{random adjacency list} of RANTEXT is generally larger than the \textit{static adjacency list} in CUSTEXT+ and smaller than that in SANTEXT+, which provides a better balance between utility and
privacy protection of perturbation.

\subsection{Trade-off between Privacy and Utility}\label{sec:trade-off}
In this subsection, we compare RANTEXT with CUSTEXT+ and SANTEXT+ in terms of privacy-utility trade-offs. we conduct experiments on the CNN/Daily Mail dataset using Vicuna-7b-4bit(3.89GB) as the extraction module. As shown in~\autoref{fig:trade-off}, each point represents the privacy protection level (under top-1 embedding inversion attack) and generation quality of a specific perturbation mechanism and a $\epsilon$ value. The yellow straight line indicates the generation quality of directly using GPT-4 without any privacy protection, referred to as `\textit{non-private}'. The experimental results demonstrate that RANTEXT tends to offer the best generation quality under the same privacy protection level compared to baseline methods. Due to the effectiveness of extraction, the coherence of
\Name is higher than that of non-private GPT-4 in most cases. 

Furthermore, our proposed method works effectively in commercial~\cite{sharma2022finred} and medical~\cite{jin2021disease} domains. More detailed experiments can be found in Appendix E and Appendix F.

In summary, RANTEXT demonstrates superior privacy protection against various attacks on differentially private mechanisms compared to baselines, confirming its robust privacy safeguarding alongside high-quality text generation.

\section{Discussion and Limitations}
\subsection{Performance Gap in MAUVE}
% Although experimental results demonstrate the effectiveness of \Name in privacy-preserving text generation, a notable gap in MAUVE scores persists when it was compared to GPT-4, as shown in \autoref{tbl:diff_size}. One probable reason for this discrepancy is the semantic perturbations introduced by LDP, which disturbs the original information in the raw prompt. Future work focused on proposing a new differentially private mechanism with a better utility-privacy trade-off could improve the generation quality in the MAUVE metric.

% Furthermore, it is important to note that the local model of the extraction module in \Name is pre-trained and not specifically fine-tuned for this task. Enhancing the extraction and reconstruction capabilities of this local model could also potentially result in improved MAUVE scores.
Although experimental results demonstrate the effectiveness of \Name in privacy-preserving text generation, a notable gap in MAUVE scores persists when compared to GPT-4, as shown in \autoref{tbl:diff_size}. One probable reason for this discrepancy is the semantic perturbations introduced by LDP, which disturb the original information in the raw prompt. Future work focusing on developing a differentially private mechanism with a better trade-off between utility and privacy protection could improve the MAUVE score.

Furthermore, it is important to note that the local model within the extraction module of \Name is pre-trained and not specifically fine-tuned for this task. Future work that enhances the extraction and reconstruction capabilities of this local model could also result in improved MAUVE scores.
%Moreover, proposing a new differentially private mechanism with a better utility-privacy trade-off could also lead to enhanced generation results.

\subsection{Comparing to Prompt Engineering Methods}
% The prompt engineering method represented by HaS~\cite{chen2023hide} trains two language models to identify private entities and randomly replace them with new words for classification and translation tasks. However, it is not implemented for open-ended text generation tasks, which are discussed in this paper. Unlike \Name that perturbs raw words with ones close in embedding distance, the perturbation in HaS is required to be semantically random. As previously mentioned, even a slight semantic bias could lead to different results in text generation tasks. The unconstrained perturbation in HaS likely makes it unsuitable for text generation tasks. Furthermore, HaS only protects specific words, leaving others exposed to adversaries.
The prompt engineering method, represented by HaS~\cite{chen2023hide}, trains a language model to identify private entities and randomly replace them with new words sampled by another language model. Experimental results demonstrate its effectiveness for privacy-preserving classification and translation tasks. However, the perturbed text in HaS is not required to be semantically relevant to the raw text. The unconstrained perturbation makes it unsuitable for open-ended text generation tasks, as its significant semantic bias could lead to semantically irrelevant generations. Furthermore, it only protects specific words of private entities, leaving others (that are not detected by HaS) exposed to adversaries.

DP-Prompt \cite{DBLP:conf/emnlp/UtpalaHC23} leverages the power of pre-trained LLMs and zero-shot prompting to counter author de-anonymization attacks \cite{bevendorff2019heuristic} while minimizing the impact on downstream utility. It provides LDP-based privacy protection for classification tasks specifically against de-anonymization attacks. However, the demo\footnote{\url{https://github.com/SaitejaUtpala/dp_prompt/blob/main/data/chatgpt_data/chatgpt_zero_shot_paraphrase_imdb.zip}} of DP-Prompt reveals the privacy leakage of personally identifiable information (PII) from its sentence-level perturbations. More importantly, it does not address the information bias introduced by the LDP, thus rendering it unsuitable for text generation tasks.

Compared to existing HaS, our proposed \Name utilizes LDP to replace the raw token with a randomly selected one that is close in embedding distance, thereby maintaining the utility of the perturbed text. To address the semantic bias that DP-Prompt does not solve, \Name locally deploys a small language model to generate an aligned output with the input of the perturbed generation and the raw document.
\subsection{Limitations of InferDPT}
The framework for privacy-preserving text generation presented in this paper has two main limitations. 

First, \Name requires the deployment of a small language model in its extraction module. In scenarios with extremely limited computational resources (e.g., smartwatch \cite{chen2014duet}), this requirement might not be feasible.

Second, there exists a gap in MAUVE scores between \Name and direct usage of GPT-4. Future research could focus on enhancing the extraction and reconstruction capabilities of the local model, for example, by optimizing the system prompt \cite{wen2024hard} of this extraction model. Additionally, the token perturbation in \Name is not optimal \cite{geng2015optimal}. Developing an optimal perturbation mechanism \cite{geng2015optimal} for text could further improve the MAUVE score.
% Second, \Name introduces additional computation for generating the perturbed document and for inferring the local model. There is still room for developing a more time-efficient framework for privacy-preserving text generation.

\subsection{Privacy Budget of InferDPT}
As previously mentioned, the perturbation module of \Name generates a perturbed document by replacing each token in the raw document with a new one sampled using local differential privacy (LDP). It is important to note that each token in the raw document undergoes the LDP process only once. Therefore, the token-level perturbation~\cite{yue-etal-2021-differential,chen2023customized} in \Name introduces no accumulated privacy risks. For instance, when using RANTEXT (which satisfies $\epsilon$-LDP for its sampling process) as the perturbation module, the privacy budget for each raw token in \Name remains $\epsilon$.

\section{Related Work}
% CipherGPT~\cite{hou2023ciphergpt} has applied homomorphic encryption~\cite{gentry2009fully} to language models that are based on Transformer. It performs inference on encrypted data. However, it results in a problem that cannot be completely solved today: the significant computation time and communication costs: it infers a token costing 24 minutes and 93 GB of bandwidth, making deployment impractical.~PromptPATE~\cite{duan2023flocks} and DP-OPT~\cite{hong2023dp} have utilized differential privacy (DP) to reconstruct the datasets used for classification tasks, thereby protecting the privacy of training data during the prompt learning (tuning) process. However, they focus on the classification tasks and do not solve the information distortion introduced by the differential privacy noise. Tang et al.~\cite{tang2023privacy} introduced a differentially private approach to generate privacy-preserving examples for in-context learning. They deploy a large language model to reconstruct the private examples via a few-shot generation of differential privacy. They also focus on classification tasks.\\
\mypara{Secure two-party inference} Iron~\cite{hao2022iron} and CipherGPT~\cite{hou2023ciphergpt} have applied homomorphic encryption~\cite{gentry2009fully} to language models that are based on Transformer~\cite{vaswani2017attention}. They perform inference on encrypted data. However, it results in a problem that cannot be completely solved today: the significant computation time and communication costs. Taking CipherGPT as an example, it infers a token costing 24 minutes and 93 GB of bandwidth, making the deployment of encrypted inference impractical.

\mypara{Privacy-preserving prompt learning (tuning)} PromptPATE~\cite{duan2023flocks} and DP-OPT~\cite{hong2023dp} have utilized differential privacy (DP) to reconstruct the datasets used for classification tasks, thereby protecting the privacy of training data during the prompt learning (tuning) process~\cite{ding2021openprompt}. However, these methods do not protect the private data of users in the prompt during the inference process with LLMs. Also, they focus on the classification tasks and do not solve the information distortion introduced by the noise of differentially private mechanisms.

\mypara{Privacy-preserving in-context learning} Tang et al.~\cite{tang2023privacy} introduced a differentially private approach to generate privacy-preserving examples for in-context learning~\cite{brown2020language}. They deploy a large language model to reconstruct the private examples via the few-shot generation of differential privacy. They also focus on the classification task and do not protect the input document during the inference process of generation tasks.

\mypara{Privacy-preserving model training} SANTEXT+~\cite{yue-etal-2021-differential} and CUSTEXT+~\cite{chen2023customized} have utilized differential privacy to enhance text privacy. They sequentially substitute words in texts with semantically similar words to preserve privacy during training in classification tasks. These two mechanisms are resistant to the input inference attack~\cite{yue-etal-2021-differential}. However, they are vulnerable to the embedding inversion attacks~\cite{qu2021natural}. They do not solve semantic distortion caused by DP noise. They are unsuitable for direct use in text generation tasks.

\section{Conclusion}
This paper explores the challenge of privacy leakage in text generation tasks executed by black-box large language models and introduces \Name as a potential solution. Additionally, we propose RANTEXT, a novel differential privacy algorithm designed for large language models following the exponential mechanism to enhance user privacy protection. We expect that our solution and findings can provide technical insights into the current privacy challenges and shed light on potential future explorations in privacy protection within emerging LLMs. 
\bibliographystyle{ieeetr}
\bibliography{jobname}


\appendix
\section*{A. Extraction Module Prompt}\label{app:B}
{The prompt for the extraction module is as follows:} \\
Your task is to extend the \emph{``Prefix Text''}. Use the \emph{``Perturbed Generation''} as your primary writing material for your extension. Extract coherent and consistent text from the \emph{``Perturbed Generation''} and integrate them into your continuation. Ensure a seamless alignment with the context established by the \emph{``Prefix Text''}. Provide only your ``Extended Text''\\
——\emph{``Prefix Text''}:\\
——\emph{``Perturbed Generation''}:\\
——\emph{``Extended Text''}:~.

\section*{B. Random Function}\label{sec:noise}
In our study, we noted that generating adjacency lists directly with Laplace distribution led to excessively large sizes. To tackle this, we created an adjusted random vector by {$cure\!\_fit(\cdot)$}\footnote{SciPy Homepage: https://scipy.org}, aiming to achieve specific probability targets for the ratio,  \( \frac{\text{Size of } C_r}{\text{Size of } V} \) of the token \textit{happy}, equaling 5\% at different \( \epsilon \) values:
\begin{table}[h]
\centering
\caption{Desired probabilities under different \( \epsilon \).}
\begin{tabular}{|c|c|}
\hline
\(\epsilon\) & Probability When \( \frac{\text{Size of } C_r}{\text{Size of } V} = 0.05 \) \\ \hline
2.0          & 1.5\%                                                           \\ \hline
6.0          & 9.0\%                                                           \\ \hline
10.0         & 10.0\%                                                          \\ \hline
14.0         & 10.5\%                                                          \\ \hline
\end{tabular}
\end{table}
\begin{equation*}
Y \sim f(x)=  \frac{Z}{2\Delta{\phi}}\cdot\exp\left(-\frac{Z\cdot|x |}{\Delta{\phi}}\right),\\
\end{equation*}
\begin{equation*}
Z = 
\begin{cases}
  ~~{\epsilon} & \text{if}~\epsilon<2,\\
  {a \log(b \cdot \epsilon + c)+d} & \text{otherwise},
 \end{cases}\\
\end{equation*}
$\text{where}\;\Delta{\phi}\;\text{is}\;\text{the}\;\text{sensitivity}\;\text{of}\;\text{function}\;\phi(\cdot).$

\section*{C. Proofs of Theorems}\label{app:proof}

\begin{proof}[\bf Proof of Theorem 1] 
Given that the output of Laplace distribution spans the range $(-\infty, \infty)$, it can be deduced that:
\begin{equation}\label{eqdp2}
Y \in (-\infty, \infty).
\end{equation}
With~\autoref{eqdp2}, it can be further deduced that:
\begin{equation}\label{pb_d}
d_e(\hat{\phi}(t), \phi(t)) \in (0, \infty)
\end{equation}
There exists a random embedding $\hat{\phi}(t)$, satisfying the condition:
\begin{equation}
d_e(\hat{\phi}(t), \phi(t)) > d_e(\phi(t'), \phi(t))
\end{equation}
Consequently, a random adjacency list $C\!_r(t)$ of RANTEXT can be constructed, fulfilling the condition $t' \in C\!_r(t)$.\\
It completes the proof.
\end{proof} 

\begin{proof}[\bf Proof of Theorem 2] 

Given a privacy parameter \(\epsilon \geq 0\) and a random adjacency list \(C\!_r(t)\) of token \(t\), for any two input tokens \( x, x' \in  C\!_r(t) \) and output token \( y \in C\!_r(t) \), it holds that:

\begin{align}
        \displaystyle \frac{\displaystyle Pr[y|x]}{\displaystyle Pr[y|x']}= \frac{\exp\left(\frac{\epsilon \cdot u(x,y)}{2\Delta u}\right)}{\sum_{y'\in C\!_r(t)} \exp\left(\frac{\epsilon \cdot u(x,y')}{2\Delta u}\right)}/\frac{\exp\left(\frac{\epsilon \cdot u(x',y)}{2\Delta u}\right)}{\sum_{y'\in C\!_r(t)} \exp\left(\frac{\epsilon \cdot u(x',y')}{2\Delta u}\right)}
\end{align}
With~$\Delta u=1$, it can be further deduced that:
\begin{equation}
            \displaystyle \frac{\displaystyle Pr[y|x]}{\displaystyle Pr[y|x']}=\frac{\exp\left(\frac{\epsilon \cdot u(x,y)}{2}\right)}{\exp\left(\frac{\epsilon \cdot u(x',y)}{2}\right)}\cdot \frac{\sum_{y'\in C\!_r(t)} \exp\left(\frac{\epsilon \cdot u(x',y')}{2}\right)}{\sum_{y'\in C\!_r(t)} \exp\left(\frac{\epsilon \cdot u(x,y')}{2}\right)}
\end{equation}
With~$0~<u(x,y)\leq~1$, it can be further deduced that:
\begin{equation}\label{ll1}
   \frac{\exp\left(\frac{\epsilon \cdot u(x,y)}{2}\right)}{\exp\left(\frac{\epsilon \cdot u(x',y)}{2}\right)} \leq \exp\left(\frac{\epsilon}{2}\right)
\end{equation}
\begin{equation}\label{eqll}
   {\exp\left(\frac{\epsilon \cdot u(x',y')}{2}\right)} \leq \exp\left(\frac{\epsilon}{2}\right)\cdot\exp\left(\frac{\epsilon \cdot u(x,y')}{2}\right)
\end{equation}
With~\autoref{eqll}, it can be further deduced that:
\begin{equation}\label{ll2}
   \frac{\sum_{y'\in C\!_r(t)} \exp\left(\frac{\epsilon \cdot u(x',y')}{2}\right)}{\sum_{y'\in C\!_r(t)} \exp\left(\frac{\epsilon \cdot u(x,y')}{2}\right)} \leq \exp\left(\frac{\epsilon}{2}\right)
\end{equation}
With~\autoref{ll1} and~\autoref{ll2}, it can be deduced that:
\begin{align}
    &\frac{Pr[y|x]}{Pr[y|x']} \leq \exp\left(\frac{\epsilon}{2}\right)\cdot  \exp\left(\frac{\epsilon}{2}\right)=e^{\epsilon}
\end{align}
It completes the proof.
\end{proof}

\begin{table*}[t]
\centering
\caption{Privacy protection levels\,$\uparrow$ of SANTEXT+, CUSTEXT+, and RANTEXT on the FinRED Dataset.}
\label{tbl:more_att}
\resizebox{0.93\textwidth}{!}{
\begin{tabular}{l|l||c c c|c c c|c c c}
\toprule
\multirow{2}{*}{\textbf{Dataset}} & \multirow{2}{*}{\textbf{Method}} & \multicolumn{3}{c|}{{BERT Inference Attack}} & \multicolumn{3}{c|}{{Embedding Inversion Attack}} & \multicolumn{3}{c}{{GPT Inference Attack}} \\ 
\cline{3-11}
\rule{0pt}{2.3ex} & & $\epsilon=2.0$ & $~\epsilon=6.0$ & $\epsilon=10.0$ & $\epsilon=2.0$ & $~\epsilon=6.0$ & $\epsilon=10.0$ & $\epsilon=2.0$ & $~\epsilon=6.0$ & $\epsilon=10.0$ \\ 
\cline{1-11}
\rule{0pt}{2.8ex}
\multirow{3}{*}{\textbf{FinRED}} & SANTEXT$^+$ &$0.831$&$0.775$ & $0.767$ &$0.631$ &$0.601$ & $0.597$  &$0.619$ & $0.594$ &$0.589$ \\
& CUSTEXT$^+$ &$0.865$ & $0.780$ &$0.678$ &$0.490$ & $0.301$ &$0.093$ &$ 0.598$ & $0.431$ &$0.250$ \\
& RANTEXT &$\mathbf{0.991}$ & $\mathbf{0.987}$ &$\mathbf{0.973}$ &$\mathbf{0.944}$ & $\mathbf{0.911}$ &$\mathbf{0.804}$ &$\mathbf{0.964}$ &$\mathbf{0.901}$  &$\mathbf{0.784}$ \\
\bottomrule
\end{tabular}
}
\end{table*}

\begin{table*}[t]
\centering
\caption{Performance comparison of different methods about open-ended text generation tasks on the FinRED Dataset.}
\label{tab:more-result}
\resizebox{0.93\textwidth}{!}{
\begin{tabular}{l|l||c c c|c c c|c c c}
\toprule
\multirow{2}{*}{\textbf{Dataset}} & \multirow{2}{*}{\textbf{Method}}  & \multicolumn{3}{c|}{\textbf{diversity$\uparrow $ }} & \multicolumn{3}{c|}{\textbf{MAUVE$\uparrow $ }} & \multicolumn{3}{c}{\textbf{coherence$\uparrow $ }} \\ 
\cline{3-11}
   \rule{0pt}{2.3ex}                      &                          & $\epsilon=2.0$ & $\epsilon=6.0$ & $\epsilon=10.0$ & $\epsilon=2.0$ & $\epsilon=6.0$ & $\epsilon=10.0$ & $\epsilon=2.0$ & $\epsilon=6.0$& $\epsilon=10.0$  \\ 
\cline{1-11}
 \rule{0pt}{2.5ex}
\multirow{6}{*}{\textbf{FinRED}} & GPT-4  &&{$0.989$} & &&{$0.284$} & &&{$0.676$} \\
& Vicuna-7b-4bit (3.89GB) &&{$0.953$} & &&{$0.106$} & &&{$0.526$} \\
\cline{2-11}
 \rule{0pt}{2.5ex}
& \Name+ SANTEXT$^+$ & $0.979$ & $0.980$ & $0.977$ & $0.118$ & $0.141$ & $0.156$ & $0.687$ & $0.690$ & $0.713$ \\
& \Name+ CUSTEXT$^+$ & $0.976$ & $0.975$ & $0.977$ & $\mathbf{0.204}$ & $\mathbf{0.244}$ & $\mathbf{0.257}$ & $0.677$ & $0.683$ & $0.703$ \\
& \Name+ RANTEXT & $\mathbf{0.983}$ & $\mathbf{0.983}$ & $\mathbf{0.984}$ & ${0.132}$ & $0.152$ & ${0.184}$ & $\mathbf{0.691}$ & $\mathbf{0.703}$ & $\mathbf{0.725}$ \\
\bottomrule
\end{tabular}
}
\end{table*}

\section*{D. Prompt for GPT inference attack}\label{app:C}
The prompt of GPT inference attack is as follows:

\textit{Given a list of tokens, denoted as ``INPUTS,'' which have been subjected to text perturbation through differential privacy techniques, your task is to recover each token in the list to its most likely original one.} \\
\textit{As an example, consider the list ``INPUTS'': \\ \text{[}"Privacy", "LLM", "Text"\text{]}. \\
The expected output format should be:} \\
\textit{[} \\
\textit{["Prediction"], \# Corresponding to "Privacy"} \\
\textit{["Prediction"], \# Corresponding to "LLM"} \\
\textit{["Prediction"] \# Corresponding to "Text"} \\
\textit{]} \\
\textit{"Prediction" represents the most plausible original tokens before perturbation.} \\
\textit{For the given list of ``INPUTS'':} \\
\textit{[INPUT HERE]} \\
\textit{Generate predictions for each token in the list.}

\section*{E. Experiments on the commercial dataset}
We also study privacy-preserving inference in the financial domain. We implement \Name on the FinRED~[23] dataset, which consists of earnings call transcripts and financial news articles. We evaluate its utility and privacy protection levels in open-ended text generation tasks.

\autoref{tbl:more_att} shows the privacy protection levels of various differentially private mechanisms against several attacks, including the BERT Inference Attack, the (top-1) Embedding Inversion Attack, and the GPT Inference Attack. Experimental results indicate that RANTEXT provides better privacy protection when compared to SANTEXT+ and CUSTEXT+ at the same privacy budget. Specifically, the privacy protection level of RANTEXT achieves a $7.65\times$ improvement over CUSTEXT+ at a privacy parameter of $\epsilon=10.0$ against the Embedding Inversion Attack. The robust
privacy safeguarding of RANTEXT benefits from its special designs of the \textit{random adjacency list} (generally larger than that in CUSTEXT+).

\autoref{tab:more-result} shows the quality of text generated by \Name with various differentially private mechanisms on the FinRED dataset. It is observed that the quality of text generated by \Name is
comparable to that directly produced by non-private GPT-4 and better than that directly produced by the local model. It proves that \Name works effectively in the financial domain. In terms of diversity and coherence, the quality of text generated by RANTEXT is superior to that of CUSTEXT+ and SANTEXT+. However, the MAUVE score of RANTEXT is inferior to that of CUSTEXT+. This is probably due to that RANTEXT discards financial nouns (those not belonging to $V$) for privacy protection during its perturbation. CUSTEXT+ keeps all of these sensitive tokens without perturbation, which results in privacy leakage.
We emphasize that this is also one of the reasons why the CUSTEXT+ is vulnerable to the embedding inversion attack.

\begin{table*}[t]
\centering
\caption{Privacy protection levels\,$\uparrow$ of SANTEXT+, CUSTEXT+, and RANTEXT on the MedQA dataset.}
\label{tbl:more_att-2}
\resizebox{0.93\textwidth}{!}{
\begin{tabular}{l|l||c c c|c c c|c c c}
\toprule
\multirow{2}{*}{\textbf{Dataset}} & \multirow{2}{*}{\textbf{Method}} & \multicolumn{3}{c|}{{BERT Inference Attack}} & \multicolumn{3}{c|}{{Embedding Inversion Attack}} & \multicolumn{3}{c}{{GPT Inference Attack}} \\ 
\cline{3-11}
\rule{0pt}{2.3ex} & & $\epsilon=2.0$ & $~\epsilon=6.0$ & $\epsilon=10.0$ & $\epsilon=2.0$ & $~\epsilon=6.0$ & $\epsilon=10.0$ & $\epsilon=2.0$ & $~\epsilon=6.0$ & $\epsilon=10.0$ \\ 
\cline{1-11}
\rule{0pt}{2.8ex}
\multirow{3}{*}{\textbf{MedQA}} & SANTEXT$^+$ &$0.831$&$0.793$ & $0.785$ &$0.634$ &$0.612$ & $0.610$  &$0.628$ & $0.608$ &$0.598$ \\
& CUSTEXT$^+$ &$0.831$ & $0.733$ &$0.583$ &$0.453$ & $0.276$ &$0.091$ &$ 0.548$ & $0.397$ &$0.246$ \\
& RANTEXT &$\mathbf{0.960}$ & $\mathbf{0.933}$ &$\mathbf{0.911}$ &$\mathbf{0.932}$ & $\mathbf{0.901}$ &$\mathbf{0.823}$ &$\mathbf{0.959}$ &$\mathbf{0.917}$  &$\mathbf{0.878}$ \\
\bottomrule
\end{tabular}
}
\end{table*}

\begin{table*}[t]
\centering
\caption{Performance comparison of different methods about open-ended text generation tasks on the MedQA dataset.}
\label{tab:more-result-2}
\resizebox{0.93\textwidth}{!}{
\begin{tabular}{l|l||c c c|c c c|c c c}
\toprule
\multirow{2}{*}{\textbf{Dataset}} & \multirow{2}{*}{\textbf{Method}}  & \multicolumn{3}{c|}{\textbf{diversity$\uparrow $ }} & \multicolumn{3}{c|}{\textbf{MAUVE$\uparrow $ }} & \multicolumn{3}{c}{\textbf{coherence$\uparrow $ }} \\ 
\cline{3-11}
   \rule{0pt}{2.3ex}                      &                          & $\epsilon=2.0$ & $\epsilon=6.0$ & $\epsilon=10.0$ & $\epsilon=2.0$ & $\epsilon=6.0$ & $\epsilon=10.0$ & $\epsilon=2.0$ & $\epsilon=6.0$& $\epsilon=10.0$  \\ 
\cline{1-11}
 \rule{0pt}{2.5ex}
\multirow{6}{*}{\textbf{MedQA}} & Claude-3.5-haiku  &&{$0.942$} & &&{$0.733$} & &&{$0.753$} \\
& Llama3.1-8b-4bit (4.9GB) &&{$0.678$} & &&{$0.562$} & &&{$0.550$} \\
\cline{2-11}
 \rule{0pt}{2.5ex}
& \Name+ SANTEXT$^+$ & $0.937$ & $0.936$ & $0.933$ & $0.562$ & $0.576$ & $0.582$ & $0.525$ & $0.552$ & $0.576$ \\
& \Name+ CUSTEXT$^+$ & $0.923$ & $0.925$ & $0.920$ & ${0.631}$ & $\mathbf{0.649}$ & ${0.656}$ & $0.641$ & $0.668$ & $\mathbf{0.690}$ \\
& \Name+ RANTEXT & $\mathbf{0.938}$ & $\mathbf{0.939}$ & $\mathbf{0.944}$ & $\mathbf{0.633}$ & $0.643$ & $\mathbf{0.664}$ & $\mathbf{0.661}$ & $\mathbf{0.676}$ & $0.680$ \\
\bottomrule
\end{tabular}
}
\end{table*}
\section*{F. Experiments on the medical dataset}

We further study the privacy-preserving inference with Claude-3.5-haiku~[8] in the medical domain. We implement \Name on the MedQA~[32] dataset, which comprises medical text questions and corresponding answers. We utilize Llama3.1-8b~[20] as the local model in the extraction module. We evaluate its utility and privacy protection levels in the open-ended text generation tasks.

\autoref{tbl:more_att-2} shows the privacy protection levels of various differentially private mechanisms against several attacks, including the BERT Inference Attack, the (top-1) Embedding Inversion Attack, and the GPT Inference Attack. Experimental results indicate that RANTEXT provides better privacy protection when compared to SANTEXT+ and CUSTEXT+ at the same privacy budget. Specifically, the privacy protection level of RANTEXT achieves a $8.03\times$ improvement over CUSTEXT+ at a privacy parameter of $\epsilon=10.0$ against the Embedding Inversion Attack. The robust
privacy safeguarding of RANTEXT benefits from its special designs of the \textit{random adjacency list} (generally larger than that in CUSTEXT+), which perturbs more raw tokens to the new ones without retaining them.

\autoref{tab:more-result-2} shows the quality of text generated by \Name with various differentially private mechanisms on the MedQA dataset. It is observed that the quality of text generated by \Name is
comparable to that directly produced by non-private Claude-3.5-haiku and better than that directly produced by the Llama3.1-8b-4bit. It proves that \Name works effectively in the medical domain. And the quality of text generated by RANTEXT and CUSTEXT+ outperforms that of SANTEXT+. This is likely because SANTEXT+ uses the entire vocabulary as its \textit{static adjacency list}, which is too large for the utility of the perturbed text.

In summary, experimental results demonstrate that our method is effective on commercial models for open-ended text generation tasks using the medical dataset.
\vfill

\end{document}